\documentclass[12pt]{article}
\usepackage{amsmath}
\usepackage{multirow}
\usepackage{amsfonts}
\usepackage{amssymb,graphics,psfrag}
\usepackage{array,epsfig,multirow,graphicx}
\usepackage{comment}
\usepackage{slashed}
\usepackage{hyperref}
\usepackage{tensor}
\usepackage{booktabs}
\usepackage{colortbl}
\usepackage{hhline}
\usepackage{mathtools}
\usepackage{enumitem}
\usepackage{xfrac}
\usepackage{tikz}
\usetikzlibrary{matrix}
\usepackage[nosort]{cite}
\usetikzlibrary{decorations.markings}
\usetikzlibrary{shapes,arrows}
\usetikzlibrary{decorations.pathreplacing}
\usetikzlibrary{arrows,positioning}

\def\hybrid{\topmargin -20pt    \oddsidemargin 0pt
        \headheight 0pt \headsep 0pt 
        \textwidth 6.25in      
        \textheight 9 in      
        \marginparwidth .875in
        \parskip 5pt plus 1pt
          \jot = 1.5ex
  }
\hybrid
\numberwithin{equation}{section}
\numberwithin{table}{section}

\newcommand{\beq}{\begin{equation}}
\newcommand{\eeq}{\end{equation}}
\newcommand{\be}{\begin{equation}}
\newcommand{\ee}{\end{equation}}
\newcommand{\bea}{\begin{eqnarray}}
\newcommand{\eea}{\end{eqnarray}}
\newcommand{\ben}{\begin{eqnarray*}}
\newcommand{\een}{\end{eqnarray*}}               
\newcommand{\ba}{\begin{aligned}}
\newcommand{\ea}{\end{aligned}}
\newcommand{\bt}{\begin{tabular}}
\newcommand{\et}{\end{tabular}}
\newcommand{\bc}{\begin{center}}
\newcommand{\ec}{\end{center}}

\newcommand{\cO}{\mathcal{O}}

\newcommand{\cC}{\mathcal{C}}
\newcommand{\cD}{\mathcal{D}}
\newcommand{\cL}{\mathcal{L}}

\newcommand{\cK}{\mathcal{K}}
\newcommand{\cN}{\mathcal{N}}

\newcommand{\cA}{\mathcal{A}}

\newcommand{\cB}{\mathcal{B}}

\newcommand{\cJ}{\mathcal{J}}
\newcommand{\cR}{\mathcal{R}}

\newcommand{\I}{\text{Im}}
\newcommand{\R}{\text{Re}}

\newcommand{\bbZ}{\mathbb{Z}}

\newcommand{\bbC}{\mathbb{C}}
\newcommand{\bbP}{\mathbb{P}}

\newcommand{\nn}{\nonumber}

\newcommand{\cref}{{\bf [check ref]}}

\definecolor{mppgreen}{RGB}{17,102,86}
\definecolor{mppgray}{RGB}{221,222,214}

\def\blfootnote{\xdef\@thefnmark{}\@footnotetext}
\long\def\symbolfootnote[#1]#2{\begingroup%
\def\thefootnote{\fnsymbol{footnote}}\footnote[#1]{#2}\endgroup}

\begin{document}

\baselineskip=15pt

\begin{titlepage}
\begin{flushright}
\parbox[t]{1.8in}{\begin{flushright}  MPP-2017-8 \end{flushright}}
\end{flushright}

\begin{center}

\vspace*{ 1.2cm}

{\Large \bf Three-form periods on Calabi-Yau fourfolds:\\[.2cm]
                  {\large Toric hypersurfaces and F-theory applications}}

\vskip 1.2cm

\renewcommand{\thefootnote}{}
\begin{center}
 {Sebastian Greiner$^{\, 1,2}$ and Thomas W.~Grimm$^{\,1,2}$\footnote{sgreiner@mpp.mpg.de,\ t.w.grimm@uu.nl}}
\end{center}
\vskip .2cm
\renewcommand{\thefootnote}{\arabic{footnote}}

{$^1$ Institute for Theoretical Physics and \\
Center for Extreme Matter and Emergent Phenomena,\\
Utrecht University, Leuvenlaan 4, 3584 CE Utrecht, The Netherlands\\[.2cm]

$^2$ Max-Planck-Institut f\"ur Physik, \\
F\"ohringer Ring 6, 80805 Munich, Germany
}

 \vspace*{.2cm}

\end{center}

 \renewcommand{\thefootnote}{\arabic{footnote}}
 
\begin{center} {\bf ABSTRACT } \end{center}

The study of the geometry of Calabi-Yau fourfolds is 
relevant for compactifications of string theory, M-theory, and F-theory 
to various dimensions. This work introduces the mathematical machinery to derive the complete 
moduli dependence of the periods of non-trivial three-forms for fourfolds realized 
as hypersurfaces in toric ambient spaces. It sets the stage to determine Picard-Fuchs-type differential equations and 
integral expressions for these forms. The key tool is the observation that non-trivial three-forms on 
fourfold hypersurfaces in toric ambient spaces always stem from divisors that are build out of trees of toric surfaces fibered over 
Riemann surfaces. The three-form periods are then non-trivially related to the one-form periods of these Riemann surfaces.
In general, the three-form
periods are known to vary holomorphically over the complex structure 
moduli space and play an important role in 
the effective actions arising in fourfold compactifications. 
We discuss two explicit example fourfolds for F-theory compactifications
in which the three-form periods determine axion decay constants. 

\vspace*{.5cm}

\hfill {February, 2017}
\end{titlepage}

\tableofcontents

\newpage



\section{Introduction}
The study of string theory compactifications on Calabi-Yau manifolds has a long tradition. This can be traced 
back to the fact that these geometries provide background solutions to all orders in $\alpha'$
that yield supersymmetric effective theories. Due to their apparent importance 
for string theory compactifications to four space-time dimensions, 
much focus has been put on the study of Calabi-Yau 
threefolds. This led to an increasingly deep understanding of the quantum 
geometry of these backgrounds and a rapid advancement of various studies
 of mirror symmetry.  
In contrast, the study of Calabi-Yau fourfolds has 
attracted much less attention. In particular, Calabi-Yau fourfolds 
can admit a non-trivial cohomology group of three-forms, whose dimension is 
neither related to the number of complex structure nor the number of K\"ahler structure deformations of 
the geometry. The new non-vanishing Hodge number on manifolds of this type is 
$h^{2,1}$ counting the non-trivial $(2,1)$-forms.
These non-trivial $(2,1)$-forms yield massless scalars or vectors 
in the effective theories obtained by compactifications on Calabi-Yau fourfolds. 
In this work we aim to study the variations of these three-forms when 
changing the moduli of the geometry. This dependence is captured 
by the periods of the three-forms, which are integrals over fixed three-dimensional 
cycles in the fourfold.

Compactifications on Calabi-Yau fourfolds lead to different effective theories 
depending on the starting point. Starting with Type IIA supergravity one finds 
a two-dimensional effective $(2,2)$-dilaton supergravity theory first studied 
in \cite{Gates:2000fj}. A complete inclusion of the three-form degrees of freedom can 
be found in \cite{Haack:2000di,Gukov:2002iq}. Using instead eleven-dimensional supergravity, the 
low energy limit of M-theory, the Calabi-Yau fourfold reduction yields a three-dimensional 
effective supergravity theory with $\cN=2$ supersymmetry \cite{Haack:2001jz,Berg:2002es}. 
If one further demands that these Calabi-Yau fourfolds are torus-fibered 
then one can find a lift of the full M-theory compactification on the fourfold
to an F-theory compactification to four dimensions \cite{Grimm:2010ks}. In other words, 
F-theory on an elliptically fibered Calabi-Yau fourfold will yield a 
four-dimensional effective supergravity theory with $\cN=1$ supersymmetry. 
In the various effective theories the three-form periods determine 
different couplings. For example, in the F-theory compactifications the 
three-form periods stemming from base three-forms determine 
the gauge coupling function of four-dimensional $\cN=1$ vector fields. The latter 
is known to be holomorphic in the moduli fields of the effective theory. In 
addition, the other three-form periods are key in the K\"ahler potential
determining the dynamics of four-dimensional $\cN=1$ complex scalar fields.
These scalar fields are naturally containing axions, i.e.~scalars with 
classical shift symmetries, as discussed in detail in \cite{Corvilain:2016kwe}. Therefore, 
the three-form periods will determine the axion decay constants and 
it is an interesting question to determine their precise value in such 
an F-theory setting \cite{Grimm:2014vva}.

It is a general fact about the variations of Hodge-structures that the 
periods of $(2,1)$-forms can be chosen to vary holomorphically in the 
complex structure moduli. Furthermore, one expects that they 
satisfy a differential equation of Picard-Fuchs type. To our knowledge, this differential 
equation has not been determined for any Calabi-Yau fourfold 
example so far. This will be one of the goals of this work. We first introduce toric 
hypersurfaces and review in detail how non-trivial three-forms arise for such 
spaces \cite{Klemm:1996ts,1998math.....12163M,Mavlyutov:2000hw}. Due to a no-go theorem for non-trivial
three-forms on hypersurfaces in toric Fano varieties, we have to use non-Fano ambient spaces
in which the anti-canonical hypersurface is only semiample. In these geometries the three-forms always stem from toric divisors that arise from Riemann 
surfaces over which compact toric surfaces are fibered. These Riemann surfaces generally will admit $(1,0)$-forms 
that then induce the $(2,1)$-forms of the 
Calabi-Yau fourfold via the so-called Gysin map \cite{0036-0279-33-2-R03,1998math.....12163M,Voisin03hodgetheory}. We will 
introduce this construction in more detail in the main text. We are able to propose residue expressions for the $(1,0)$-forms and 
then lift these to expressions for the $(2,1)$-forms. 
This leads us to a geometric approach to the three-form periods 
and Picard-Fuchs equations. 

It is important to point out that, similar to the analysis of periods on 
Calabi-Yau threefolds, specific boundary 
conditions at the large complex structure point can be found using 
mirror symmetry. This was done in ref.~\cite{Greiner:2015mdm}, where mirror symmetry 
for Type IIA string theory on Calabi-Yau fourfolds was discussed in detail.  
Recalling that mirror symmetry exchanges complex structure
and K\"ahler structure moduli of the geometry one can infer the behaviour 
of the periods at the large complex structure point by knowing the mirror behaviour 
at the large volume point. We have found in \cite{Greiner:2015mdm} that this fixes 
the periods to be constant or linear in the complex structure moduli
at the large complex structure point. Furthermore, the coefficients of these functions 
are given in terms of intersection numbers of two three-forms and one 
two-form on the Calabi-Yau fourfold. Combining the results of the paper \cite{Greiner:2015mdm}
with the findings we present below, the Picard-Fuchs equations can be solved 
explicitly for a given sufficiently simple example. 

In addition to the introduction of a period matrix, we also determine the structure of the intermediate Jacobian, an abelian variety that provides the moduli space of the three-form moduli, in terms of the toric data. On this space we calculate the 
natural positive definite bilinear form arising in compactifications on Calabi-Yau fourfolds. We clarify its dependence on the period matrix and certain intersection numbers that where already introduced in \cite{Greiner:2015mdm,Corvilain:2016kwe} and give a toric interpretation. Since the toric methods generalize the usual approach to string vacua obtained from Landau-Ginzburg orbifolds, \cite{Vafa:1989xc,Greene:1988ut,Vafa:1990mu,Intriligator:1990ua,Cecotti:1990kz}, we find again that the period matrix can be determined from a so called chiral ring and since these period matrices satisfy a local integration condition we propose the existence of a prepotential. This prepotential captures the complex structure dependence of the three-form couplings and its leading order behavior at large complex structure is determined by the above mentioned intersection numbers of its mirror, as found in \cite{Greiner:2015mdm}.

In this paper we discuss two interesting explicit examples. The first example will 
be a hypersurface in a toric ambient space with one non-trivial $(2,1)$-form that arises 
from a two-torus in a single exceptional divisor. The periods then obey a simple Picard-Fuchs 
equation that can be solved explicitly. Interestingly, the example geometry has an elliptic fibration and 
can thus be used as an F-theory background. The $(2,1)$-form yields 
a single four-dimensional complex scalar parameterizing the
zero-modes of the R-R and NS-NS two-forms on this background. In fact,
the two-torus yielding a $(2,1)$-form turns out to be the elliptic fiber over some divisor 
in the base, similar to the configuration considered in \cite{Grimm:2014vva}. 
The second example is significantly more involved, since it will admit seven $(2,1)$-forms 
that stem from a Riemann surface of genus seven. This geometry is also elliptically 
fibered and can serve as an F-theory background. In this case, however, the $(2,1)$-forms 
are corresponding to Wilson line moduli of seven-branes. The three-form periods 
for such scalars are relevant, for example, in the applications of refs.~\cite{Jockers:2004yj,Carta:2016ynn,Corvilain:2016kwe}.
We will discuss various interesting aspects of this example, but will not attempt to derive 
the Picard-Fuchs equations and periods explicitly. 

The paper is organized as follows. In \autoref{sec:threeform-general}
we first summarize some generalities about three-forms on Calabi-Yau fourfolds. 
In \autoref{sec:three-forms_hyper} we introduce the geometric framework in which one can construct explicit 
fourfold examples exhibiting a non-trivial three-form cohomology. Here we also recall the 
complex structure dependence of Riemann surfaces and derive Picard-Fuchs type equations and 
discuss the geometry of the intermediate Jacobian of the Calabi-Yau fourfold. In the final \autoref{sec:Example} we discuss 
examples for which these Picard-Fuchs equations can be evaluated explicitly. 
We also comment on the effective theories arising from compactifying F-theory 
on these example geometries.

%
%

\section{Three-forms on Calabi-Yau fourfolds} \label{sec:threeform-general}

In this section we first introduce some general facts about the moduli-dependence of three-forms 
on Calabi-Yau fourfolds. To do that we consider compact complex four-dimensional manifolds $Y_4$, 
which we demand to be Calabi-Yau fourfolds having exactly holonomy group $SU(4)$. 
For such geometries the Hodge numbers $h^{p,q}(Y_4) = \text{dim}(H^{p,q}(Y_4))$
have to satisfy various constraints. In fact, there are only three independent non-trivial Hodge numbers: $h^{1,1}(Y_4)$, $h^{3,1}(Y_4)$, 
and $h^{2,1}(Y_4)$. The significance of $h^{1,1}(Y_4)$ and $h^{3,1}(Y_4)$ is very similar to the case of a Calabi-Yau threefold. 
The number $h^{1,1}(Y_4)$ counts the allowed K\"ahler structure deformations, while the number $h^{3,1}(Y_4)$ counts the complex
structure deformations. The K\"ahler structure deformations will be denoted by $v^\Sigma$ and parametrize the 
expansion of the K\"ahler form $J$ into harmonic $(1,1)$-forms $\omega_\Sigma$ as
\beq \label{exp-J}
   J = v^\Sigma \, \omega_\Sigma \, \qquad \Sigma = 1,\ldots,h^{1,1}(Y_4)\ .
\eeq 
The complex structure deformations will be denoted by 
\beq
     z^\cK  , \qquad \cK=1,\ldots, h^{3,1}(Y_4)
\eeq 
in the following. It is well-known that both sets of deformations become moduli fields in the effective theory obtained by dimensional reduction of string theory, M-theory, or F-theory on $Y_4$. The Hodge number $h^{2,1}(Y_4)$ has no threefold analog. 
In fourfold compactifications of M-theory or Type IIA string theory this Hodge number counts additional complex scalars
\beq \label{N_cA-def}
    N_\cA \, , \qquad \cA = 1,\ldots,h^{2,1}(Y_4) \, ,
\eeq
that arise from the expansion of the higher-dimensional three-form into $(2,1)$-forms of $Y_4$.
Deriving 
the moduli-dependence of these $(2,1)$-forms is the main interest of this work.  
 
It is crucial to point out that a Calabi-Yau fourfold $Y_4$ with exact $SU(4)$ holonomy has $h^{3,0}(Y_4)=0$.
A general fact known from Hodge theory \cite{Voisin03hodgetheory} then implies that the $(2,1)$-forms on $Y_4$ vary holomorphically and without obstructions with the complex structure moduli $z^K$. Therefore, we can describe the variation of a $ (2,1) $-form as sections of a bundle over the complex structure moduli space with fibers parameterized by the $ (2,1) $-forms. Each fiber defines a complex $h^{2,1}$-dimensional subspace in 
the $2h^{2,1}$-dimensional cohomology group $H^3(Y_4,\mathbb{C})$. Note that we can introduce a 
real basis $(\tilde \alpha_\cA,\tilde \beta^\cB)$, $\cA,\cB=1,\ldots,h^{2,1}(Y_3)$ of $H^3(Y_4,\mathbb{R})$ such that the $(2,1)$-forms $ \psi_\cA $ are expanded as
\beq
   \psi_\cA = \Pi^\cB_\cA(z) \alpha_\cB + \tilde \Pi_{\cA \cB}(z) \beta^\cB  \, , \qquad  \Pi^\cB_\cA =\int_{A_\cB}  \psi_\cA \, ,\quad 
 \tilde  \Pi_{\cA \cB} =\int_{B^\cB}  \psi_\cA \, ,
\eeq
where $\Pi^\cB_\cA$, $\tilde \Pi_{\cA \cB}$ are the periods of $\Psi_\cA$ and vary holomorphically in the 
complex structure moduli $z^K$. The three-cycles $(A_\cA,B^\cA)$ are chosen to integrate 
to $(\delta^\cA_{\cB},\delta^\cB_{\cA})$ on $(\tilde \alpha_\cB,\tilde \beta^\cB)$, respectively, and zero otherwise. 
At this point, the split into $\tilde \alpha_\cA$ and $\tilde \beta^\cA$ is 
purely artificial, since the total space $H^3(Y_4,\mathbb{C})$ is independent of the complex structure. 
However, we can define an induced complex structure $ \cJ $ on $ H^3(Y_4,\mathbb{C}) $ 
that varies with the complex structure of the Calabi-Yau fourfold. $ \cJ $ will be 
defined to have $(2,1)$-forms in its $-i$ eigenspace and $(1,2)$-forms in its $+i$ eigenspace.

At a fixed complex structure $z_0$ the map $\cJ$ is a real endomorphism that squares to the negative identity. 
Thus, we can find around $ z_0 $ a specific real basis $( \alpha_\cA,  \beta^\cA)$ of $ H^3(Y_4,\mathbb{R}) $ 
such that
\begin{equation}
 \cJ(z_0) \, \left(\!\!\begin{array}{c}  \alpha_\cA \\  \beta^\cB \end{array} \!\!\! \right) = \left(\! \! \begin{array}{c}  \beta^\cA \\ - \alpha_\cB \end{array} \!\!\! \right)\,  .
\end{equation}
Writing a $(2,1)$-form on the Calabi-Yau fourfold at $ z_0 $ in complex structure moduli space as $\psi_\cA (z_0) = \alpha_\cA + i \beta^\cA$ 
we indeed have  $\cJ ( \psi_\cA)  = -i  \psi_\cA $.
Then there exists (locally) a holomorphic $ H^3(Y_4,\mathbb{C}) $-endormorphism-valued function $ f $, such that we can write
\begin{equation} \label{exp-psicA}
  \psi_\cA (z) = \alpha_\cA + i f_{\cA \cB} (z) \beta^\cB \quad \in H^{2,1}((Y_4)_z)
\end{equation}
to describe the local variation of a $(2,1)$-form around the point $ z_0 $. Since $ f_{\cA \cB}(z_0) = \delta_{\cA \cB}$, 
its real part is locally invertible. Denoting the inverse by $\text{Re} f ^{\cA \cB} \equiv (Re (f_{\cA \cB}))^{-1} $ we can normalize
\begin{equation} \label{ansatz_Psi}
 \Psi^\cA (z,\bar z) =\frac{1}{2} \, \text{Re}f^{\cA \cB} \big(\alpha_\cB - i \bar f_{\cB \cC} (\bar z) \beta^\cC \big) \quad \in H^{1,2}((Y_4)_z) \, .
\end{equation}
which justifies the ansatz for $(1,2)$-forms used in \cite{Grimm:2010ks,Greiner:2015mdm,Corvilain:2016kwe}. The normalized form \eqref{ansatz_Psi}
will not be of big relevance in this work, but turned out to be key in determining the effective actions 
obtained by compactification on $Y_4$. As mentioned above, the effective actions will contain new moduli fields 
$N_\cA$ arising  from the $(1,2)$-forms that parameterize the torus $H^{1,2}(Y_4)/H^{3}(Y_4,\mathbb{Z})$ 
\cite{Grimm:2010ks,Greiner:2015mdm,Corvilain:2016kwe}. It will later be convenient
to work with the holomorphic forms \eqref{exp-psicA} instead of \eqref{ansatz_Psi}. These 
forms parameterize the torus 
\beq \label{def-cJ3}
   \cJ^3 (Y_4) = \frac{H^{2,1}(Y_4)}{H^{3}(Y_4,\mathbb{Z})}\ ,
\eeq 
a space that is also known as the \textit{intermediate Jacobian} of the Calabi-Yau 
fourfold $Y_4$.

The goal of this work is to compute the periods $(\Pi^\cA_\cB(z), \tilde \Pi_{\cB \cA}(z))$ and the 
function $f_{\cA \cB}(z)$. In an appropriate basis they are related by 
\beq
   f_{\cA \cB}(z) =   (\Pi^\cA_\cC)^{-1} \tilde \Pi_{\cC \cB}\ .
\eeq
Note that from variations of Hodge structures under changes of complex structure 
one deduces that $H^{2,1}(Y_4)$ varies into $H^{1,2}(Y_4)$. 
Since $H^{0,3}(Y_4)$ is trivial, the latter varies again into $H^{2,1}(Y_4)$, such that we expect that 
$(2,1)$-forms satisfy a second order differential equation. For the considered class of geometries we 
will describe how this differential equation is determined. 

As pointed out around \eqref{N_cA-def} the non-trivial three-forms yield complex
scalar fields $N_\cA$ in the effective actions of M-theory and Type IIA string theory. Their kinetic terms are determined by an integral proportional to \footnote{See \cite{Corvilain:2016kwe} for 
a derivation of this result using the same notation and conventions.}
\beq \label{def-Q}
  Q(\Psi^\cA,\bar \Psi^\cB) \equiv \int_{Y_4} \Psi^\cA \wedge \ast \bar \Psi^\cB =  i v^\Sigma \int_{Y_4} \omega_\Sigma \wedge \Psi^\cA \wedge \bar \Psi^\cB \ ,
\eeq
where $\ast$ is the Hodge star on $Y_4$ and we have used that for a $(1,2)$-form one has $\ast \Psi^\cA = - iJ \wedge \Psi^\cA$ 
with $J$ expanded as in \eqref{exp-J}. Note that we can expand this expression further by inserting \eqref{ansatz_Psi}. 
Using the topological couplings 
\beq \label{def-MM}
  M_{\Sigma \cA}{}^{\cB} = \int_{Y_4} \omega_\Sigma \wedge \alpha_\cA \wedge \beta^\cB\ ,\qquad M_{\Sigma}{}^{\cA \cB} = \int_{Y_4} \omega_\Sigma \wedge \beta^\cA \wedge \beta^\cB\ , 
\eeq
we find
\beq \label{Q-expand}
   Q(\Psi^\cA,\bar \Psi^\cB) = -\frac{1}{2} \, \text{Re} f^{\cB \cC}  \, v^\Sigma (M_{\Sigma \cC}{}^\cA + i f_{\cC \cD} \, M_{\Sigma}{}^{\cD\cA})\ .
\eeq
When working with the holomorphic representatives \eqref{exp-psicA}, we have to multiply \eqref{Q-expand} 
with $\text{Re}f_{\cA\cB}$ appropriately, i.e.
\beq \label{Qpsipsi}
   Q(\psi_\cA,\bar \psi_\cB) = 2 \, \text{Re} f_{\cB \cC}  \, v^\Sigma (M_{\Sigma \cA}{}^\cC + i f_{\cA \cD} \, M_{\Sigma}{}^{\cD\cC})\ .
\eeq
In order to derive the metric $Q(\Psi^\cA,\bar \Psi^\cB)$ for the fields $N_\cA$ we therefore have not only to determine $f_{\cA \cB}$
as a function of the complex structure moduli $z^\cK$, but also evaluate the intersection numbers 
\eqref{def-MM} for a given geometry. In this work we will show how this can be done for Calabi-Yau fourfolds realized 
as hypersurfaces in toric ambient spaces.

\section{Three-forms on toric hypersurfaces} \label{sec:three-forms_hyper}

In this section we introduce the explicit constructions of Calabi-Yau fourfolds as hypersurfaces in toric ambient spaces. 
We explain that these spaces can admit non-trivial three-forms and that these three-forms are intimately linked 
to the existences of divisors that carry non-trivial one-forms in the Calabi-Yau geometry. These divisors are fibration 
over Riemann surfaces with fibers being toric surfaces.
The main idea is to appropriately push-forward the periods determined for the embedded Riemann surfaces 
to periods of three-forms on the fourfold. The periods of the Riemann surfaces can be derived  
by solving the associated Picard-Fuchs equations. This allows us to determine a positive definite quadratic 
form on the intermediate Jacobian introduced in the previous section in terms of the period matrices of the Riemann surfaces 
and certain intersection numbers of the ambient space. We end this section with an illustration of these 
concepts for hypersurfaces in weighted projective spaces.
In section \ref{sec:Example} we provide Calabi-Yau hypersurface examples for which 
these steps can be performed explicitly.

\subsection{Origin of non-trivial three-forms} \label{subsec:hypersurfac-general}

In this subsection we will review the generic features of the explicit construction given in \autoref{hypersurface_appendix} for smooth Calabi-Yau fourfold hypersurfaces  in toric ambient spaces. For these to be equipped with non-trivial three-form cohomology, the ambient space can not be Fano, due to the Lefschetz-hyperplane theorem and the cohomological properties of the ambient toric space.

Let us now take a look at the Lefschetz hyperplane theorem, as stated in \cite{1993alg.geom..6011B}. There it was found that for a quasi-smooth hypersurface $ Y_4 $ of a five-dimensional complete simplicial toric variety $ \cA_5 $ defined by an ample divisor that the natural map (the restriction of forms) $ \iota^\ast : H^j(\cA_5,\mathbb{C}) \rightarrow H^j(Y_4,\mathbb{C}) $ is an isomorphism for $ j \leq 3 $ and an injection for $ j = 4 $. This implies that there are no non-trivial three-forms if the divisor class of the hypersurface is ample and the hypersurface is smooth, as is the case for the sextic hypersurface in $ \mathbb{P}^5 $, since a toric variety $ \cA_5 $ does not support odd cohomology \footnote{It can be shown that for a general toric variety  $ M $, $ H^{i,j}(M) \neq 0 $ requires $ i =j $}. As a consequence, we have to deal with more complicated ambient spaces than the standard projective space to obtain non-trivial three-forms on its anti-canonical hypersurface. 

Another way to see that the ambient space $ \cA_5 $ can not be Fano to obtain a non-trivial three-form cohomology can be inferred from \cite{Mavlyutov:2000dj} where it was shown that all cohomology of degree less than four has to be induced by toric divisors $ D^\prime _l $ of $ Y_4 $. This is a set of complex codimension one submanifolds that are invariant under the toric action of the ambient space $ \cA_5 $. In particular are these toric divisors again hypersurfaces in the toric divisors  $ D_l $ of $ \cA_5 $. The precise relation as we explain in \autoref{hypersurface_appendix} is given by the so called Gysin morphism
\begin{equation} \label{one_to_three-form}
 \bigoplus \iota_{l \ast} \, : \quad \bigoplus_{\nu^\ast _l} H^1( D^\prime _l, \mathbb{C}) \longrightarrow H^3( Y_4, \mathbb{C})  \, ,
\end{equation}
where the morphism is the direct sum of Gysin morphisms $ \iota_{l \ast} $ of the inclusions $ \iota_l$ of the toric divisors $ D^\prime _l $. This is an isormophism and hence every non-trivial three-form cohomology class is a push-forward of a one-form cohomology class on a toric divisor $ D^\prime _l $.

As we show in \autoref{Divisor_appendix} not all toric divisors host a non-trivial one-form cohomology. The divisors that actually do are denoted by $ D^\prime _{l_\alpha} $ and show a fibration structure. Also, the ambient space $ D _{l_\alpha} $ of $ D^\prime _{l_\alpha} $ shows a similar fibration structure. The interesting feature here is that the base space of the $ D^\prime _{l_\alpha} $ is given by a Riemann surface $ R_\alpha $ and has fiber a toric surface $ E_{l_\alpha} $. The notation already infers the intersection properties of such divisors: if two such divisors $ D^\prime _{l_\alpha} $ intersect, they need to share the same base Riemann surface $ R_\alpha $. Therefore $ \alpha = 1, \ldots, n_2 $ counts the Riemann surfaces $ R_\alpha $ and $ l_\alpha $ counts the fibrations $ D^\prime _{l_\alpha} $ with base $ R_\alpha $ and fiber $ E_{l_\alpha} $. The Riemann surface $ R_\alpha $ is again a hypersurface in a toric ambient space $ \cA_{2,\alpha} $ which is two-dimensional and the base space of the fibration structure of $ D_{l_\alpha} $ with the same fiber $ E_{l_\alpha} $. The one-forms of the fibration $ D^\prime _{l_\alpha} $ are pull-backs of the projection $ \pi_{l_\alpha} $ to the Riemann surface $ R_\alpha $ and hence we find
\begin{equation} \label{three-cohom}
 H^{2,1}( Y_4 ) \simeq \bigoplus_{\alpha} \bigoplus_{l_\alpha} H^{1,0} (R_\alpha) \otimes H^{0,0}(E_{l_\alpha}) \, .
\end{equation}
Note that the cohomology of $ E_\alpha $ is independent of the complex structure of $ Y_4 $. The complex dimension of $ H^{1,0}(R_\alpha) $ is given by the genus $ g_\alpha $ of the Riemann surface. The spaces $ H^{1,0}(R_\alpha) $ capture correspondingly the full complex structure dependence of $ H^{2,1}(Y_4) $ which is the primary interest of this work.

Let us stress that, on the one hand, equation \eqref{one_to_three-form} implies that the non-trivial three-forms are directly inherited from the divisors $ D^\prime_{l_\alpha} $, i.e.~the divisors that are fibrations of toric surfaces $ E_{l_\alpha} $ over Riemann surfaces $ R_\alpha $ embedded in $ \cA_{2,\alpha} $. On the other hand, equation \eqref{three-cohom} indicates that 
an equivalent statement for three-forms on $D^\prime_{l_\alpha} $ cannot be made. In fact, the divisors 
$ D^\prime _{l_\alpha} $ carry in general way more non-trivial three-forms than the full Calabi-Yau fourfold $   Y_4 $, 
which, however, do not descend to $Y_4$.  

The identification \eqref{three-cohom} can also be used to infer the formula of \cite{Batyrev:1994pg,Batyrev:1994ju,Klemm:1996ts,Kreuzer:1997zg} counting the number of non-trivial $ (2,1) $-forms as
\begin{equation} \label{h21-ells}
 h^{2,1}( Y_4) = \sum_{\alpha=1}^{n_2} \ell^\prime(\theta^\ast _\alpha) \ell^\prime(\theta _\alpha) \, ,
\end{equation}
where the sum runs over pairs of dual two-dimensional faces $ (\theta^\ast _\alpha, \theta _\alpha) $. 
Recall from \eqref{introduceDl} and \eqref{fiber_D'} that $ \ell^\prime(\theta^\ast _\alpha) $ counts the divisors $ E_{l_\alpha} $ over the singular Riemann surface $ R_\alpha $.
The genus of $R_\alpha$ is given by  $ g_{\alpha} = h^{1,0}(R_\alpha) = \ell^\prime(\theta_\alpha) $. This data is only dependent on the polyhedra $ \Delta^\ast, \Delta $ and independent of the chosen triangulation.

In the following we will analyze the smooth variety $Y_4$ further and describe the complex structure variation of a $ (2,1) $-form 
on this space. We argue that this can be done by first considering the complex structure variations of $ (1,0) $-forms 
\beq \label{(1,0)forms}
    \gamma_{a_\alpha} \in H^{1,0}(R_\alpha) \, , \quad a_\alpha = 1, \ldots ,  g_\alpha \, ,
\eeq 
on $ R_\alpha $. To do so, we define holomorphic $ (1,0) $-forms on $ R_\alpha $ as Poincar\'e residues of their ambient spaces $ \cA_{2,\alpha} $. This representation for the holomorphic $ (1,0) $-forms will be explained in the next section.

\subsection{Periods of embedded Riemann surfaces and their Picard-Fuchs equations} \label{Riemann_periods}

As we have seen from the previous section, all three-forms on a Calabi-Yau fourfold hypersurface of a toric variety are induced from one-forms of Riemann surfaces. Therefore, we start this section with the basics of the theory of Riemann surfaces, as described in \cite{griffiths2011principles}. Afterwards, we restrict to the toric setting and view these Riemann surfaces as (semi-) ample hypersurfaces of a two-dimensional toric variety, as described in \cite{1993alg.geom..6011B,Mavlyutov:2000hw}. We close this section with a derivation of a second order differential equation, the Picard-Fuchs equation, that governs the complex structure dependence of the holomorphic one-forms on a Riemann surface. This is familiar from Landau-Ginzburg orbifolds as discussed in \cite{Lerche:1991wm}.

Since we are interested in the (co-)homology of the Riemann surface, a compact one-dimensional K\"ahler manifold, and the eigenspaces of its complex structure, we introduce here appropriate bases of the non-trivial cohomology groups, that allow us to perform calculations.

Consider a Riemann surface $ R $ of genus $ g $ with a basis of $ H_1(R,\mathbb{Z}) $ the one-cycles $ \hat A_a, \hat B^a $ $ a = 1, \ldots, g $ with duals $ \hat \alpha_a, \hat \beta^a \in H^1(R,\mathbb{Z})$. This basis can be chosen to be canoncial, i.e.~to satisfy
\begin{equation} \label{sympl_on_R}
 \int_{R} \hat \alpha_a \wedge \hat \beta^b = \delta_a^b \, , \quad \int_R \hat \alpha_a \wedge \hat \alpha_b = \int_R \beta^a \wedge \beta^b = 0 \, , \quad a,b = 1, \ldots, g \, .
\end{equation}

Due to a Riemann surface being K\"ahler, we can always choose a basis $ \gamma_a \in H^0(R,\Omega^1) $ of holomorphic one-forms on $ R $. Integrating these over the base of one-cycles $ \hat A_a, \hat B^a $ leads to the two period matrices $ \hat \Pi_a {} ^b, \hat \Pi_{ab} $,
\begin{equation}
 (\hat \Pi_a ) ^b = \int_{\hat A_b} \gamma_a \, , \quad ( \hat \Pi_a)_b = \int_{\hat B^b} \gamma_a \, .
\end{equation}
The periods $ \hat \Pi^b $ and $ \hat \Pi_b $ are defined to be the column vectors of these matrices, i.e. the vector formed by integrating all one-forms $ \gamma_a $ over the same one-cycle $ A_b $, $ B^b $ respectively, and
these $ 2g $ vectors are linearly independent over $ \mathbb{R} $ and hence generate the lattice
\begin{equation}
 \hat \Lambda = \bigoplus_a \big( \mathbb{Z} \hat \Pi_a \oplus \mathbb{Z} \hat \Pi^a \big)
\end{equation}
in $ \mathbb{C}^g $. This allows us to define the Jacobian variety $ \cJ^1(R) = \mathbb{C}^g  / \hat \Lambda $ of the Riemann surface $ R $ to be
\begin{equation}
 \cJ^1(R) = \frac{H^{1,0}(R)}{H^1(R,\mathbb{Z})} \simeq \mathbb{C}^g / \hat \Lambda \, .
\end{equation}

It can be shown that $ \hat \Pi_a {} ^b $ is in general invertible. We can normalize this basis to $ \tilde \gamma_a \in  H^0(R,\Omega^1)$ by multiplication with the inverse $ (\hat \Pi^{-1})_a {} ^b $ of $ \hat \Pi_a {} ^b $ such that 
\begin{equation}
 \tilde \gamma_a = (\hat \Pi^{-1})_a {} ^b \, \gamma_b \, ,  \quad \int_{A_b} \tilde \gamma_a = \delta_a^b
\end{equation}
with the remaining normalized period matrix
\begin{equation}
 i \, \hat f_{ab} = (\hat \Pi^{-1})_a {} ^c \, \hat \Pi_{cb} = \int_{B^b} \tilde \gamma_a \, .
\end{equation}
This normalized period matrix satisfies the properties
\begin{equation}
 \hat f_{ab} = \hat f_{ba} \, , \quad \R\, \hat f_{ab}\, > 0 \, .
\end{equation}
We also note that the positive definite quadratic form on $ H^0(R,\Omega^1) $ in the normalized basis is given by
\begin{equation} \label{eval_gammagamma}
- i \int \gamma_a \wedge \bar \gamma_b = 2 \cdot \R\,\hat f_{ab} \, ,
\end{equation}
where we dropped the tilde.
For our physical applications, we will be interested in complex structure dependence of the normalized period matrix $ \hat f_{ab} $ and this can be done via an explicit representation of the holomorphic one-forms $ \gamma_a $, which we will discuss in detail in \autoref{PF_appendix}.

In this appendix we give an explicit representation of the holomorphic one-forms $ \gamma_b $ on $ R $ embedded as a toric hypersurface via the Poincar\'e residue. For the full fourfold $ Y_4 $, in which we have a toric divisor a fibration with base $ R $, we find that all the $ \gamma_b \in H^{1,0}(R)$ depend only on the complex structure deformations $ a_c $, $ c=1,\ldots, h^{1,0}(R) $ of $ R $ induced by its ambient space $ Y_4 $ after blowing down the corresponding toric divisor.

As we show in \autoref{PF_appendix},
we can express the second derivatives of $ \gamma_b $ by operators acting on $ \gamma_b $ of the form
\begin{equation}
\frac{\partial}{\partial a_c} \frac{\partial}{\partial a_d} \, \gamma_b (a) = \big( c^{(1)}{(a)_{c d b e}}^ f \frac{\partial}{\partial a_e} + c^{(0)}{(a) _{c d b}}^f \big) \, \gamma_f (a) \,,
\end{equation}
where $ c^{(1)}(a) _{c d b e} \, ^ f$, $c^{(0)}(a) _{c d b} \, ^f $ are rational functions of the complex structure moduli $ a_c $ that are completely symmetric in their lower four, respectively three, indices. These functions are structure constants of the chiral ring $ \cR = \cR_\Delta $ determining the multiplication rules in this ring.
The above differential relations are called Picard-Fuchs equations and can be used to determine the complex structure dependence of the holomorphic one-forms on $ R $. In particular, this implies that the flat complex structure coordinates $ z^\cK(a) $ can still be calculated in the usual way, since these are also determined by the structure constants of $ \cR $, as described for example in \cite{Hosono:1993qy,Mayr:1996sh}. In these coordinates, we find that
\begin{equation}
 \frac{\partial}{\partial z^{\cK}}  \frac{\partial}{\partial z^{\cL}} \, \gamma_b (a(z)) = 0 \, ,
\end{equation}
which implies that $ \gamma_a (z) $ is at most linear in the $ z^\cK $ moduli. Integrating these over a basis of one-cycles we obtain the period matrices $ \hat \Pi_a {} ^b , \hat \Pi_{ab} $, which are still at most linear. This means that we can find as solutions the constant identity matrix and the normalized period matrix $ \hat f_{ab} $ that satisfies
\begin{equation}
 \hat f_{ab}(a(z)) = z^\cK \hat M_{\cK ab} + \hat C_{ab} + \cO(z^{-1}) \, ,
\end{equation}
with $ \hat M_{\cK ab}, \hat C_{ab} \in \mathbb{C} $ constants determined by boundary conditions, as was done in \cite{Greiner:2015mdm} for $ \hat M_{\cK ab} $, where it was found that these numbers arise from certain intersection numbers of the mirror Calabi-Yau fourfold, when expanding around the large complex structure point.

These considerations will be the starting point for the investigation of the intermediate Jacobian of a Calabi-Yau fourfold realized as a hypersurface in a toric variety, since in this situation all non-trivial three-form cohomology can be traced back to Riemann surfaces.

\subsection{The intermediate Jacobian of a Calabi-Yau fourfold}

In the previous subsection we have discussed the complex structure variations of the $(1,0)$-forms $\gamma_{a_\alpha}$
on the Riemann surfaces $R_\alpha$ embedded into $D_{l_\alpha}'$ and $Y_4$. Since there are
in general several such Riemann surfaces in $Y_4$ we now restore the index $\alpha$ as in \autoref{subsec:hypersurfac-general}.
In this subsection we describe how these $(1,0)$-forms are mapped to $(2,1)$-forms on $Y_4$. These forms parametrize 
the intermediate Jacobian $\cJ^3(Y_4)$ introduced in \eqref{def-cJ3} and we will describe some of its key geometrical properties. 

The precise relation between the $(1,0)$-forms $\gamma_{a_{\alpha}}$ and $(2,1)$-forms $\psi_\cA$
is inferred from the isormophism \eqref{one_to_three-form} and \eqref{three-cohom}. Explicitly it is given by
\begin{equation} \label{chi_cA_Gysin}
 \psi_\cA =  \iota_{l_\alpha \ast} \big( \pi_{l_\alpha} ^\ast \, \gamma_{a_\alpha} \big) \ , \quad \cA = ( \alpha, l_\alpha, a_\alpha ) = (1,1,1), \ldots, (n_2, \ell^\prime(\theta^\ast _\alpha),\ell^\prime(\theta_\alpha))    \, ,
\end{equation}
where we have stressed that the index $\cA$ is a multi-index labelling the Riemann surface $R_\alpha$, the toric divisors 
$D_{l_\alpha}$ that have $ R_\alpha $ as a base, and its $(1,0)$-forms $\gamma_{a_\alpha}$. The involved maps are the pullback $\pi_{l_\alpha} ^\ast$, mapping 
one-forms on $R_\alpha$ to one-forms on $D'_{l_\alpha}$, and the Gysin map $\iota_{l_\alpha \ast}$
pushing these one-forms to three-forms on $Y_4$. The Gysin map can be understood as first taking the Poincar\'e-dual 
of  $\pi_{l_\alpha} ^\ast \, \gamma_{a_\alpha}$ in $D'_{l_\alpha}$, which yields a five-cycle representing a
a homology class on $D'_{l_\alpha}$. This homology class can be pushed to the homology of 
$Y_4$ using the embedding map $\iota_{l_\beta} : D^\prime _{l_\beta} \hookrightarrow Y_4 $. Taking the Poincar\'e-dual 
of this five-homology class on $Y_4$ yields the desired three-form. As pointed out already above, the construction of $(3,2)$-forms 
$\chi_\cA$ on $Y_4$ is more straightforward, since it only involves pullbacks of the restriction morphisms. Translating \eqref{five_to_five_form}, \eqref{five-cohom} they are given by 
\begin{equation} \label{chi_cA-map}
 \chi_\cB = (\iota_{l_\beta} ^\ast )^{-1}( \omega^{(2,2)} _{l_\beta}  \wedge \pi_{l_\beta} ^\ast\, \gamma_{b_\beta} ) 
\end{equation}
where $ \omega^{(2,2)} _{l_\beta} \in H^4(D^\prime _{l_\beta},\mathbb{Z}) $ are the volume-forms of of the fibers 
$ E_{l_\beta} $ of $D^\prime _{l_\beta}$.  Note that when constructing a basis of five-forms using \eqref{chi_cA-map}, we might choose $ \omega^{(2,2)} _{l_\beta} $ topological or dependent on K\"ahler moduli. For convenience, we have chosen here the topological approach.

Let us next turn to the intermediate Jacobian $ \cJ^3(Y_4) $ spanned by the $(2,1)$-forms $\psi_\cA$. Using \eqref{one_to_three-form} we find that it splits into a direct product of Jacobians $ \cJ^1(R_\alpha) $ of Riemann surfaces $ R_\alpha $ as
\begin{equation}
 \cJ^3(Y_4) = \frac{H^{2,1}(Y_4)}{H^3(Y_4,\mathbb{Z})} \simeq \prod_{ \alpha=1}^{n_2} \big( \cJ^1(R_\alpha) \big)^{\ell^\prime (\theta^\ast _\alpha)} \, .
\end{equation}
In particular, this suggests that the period matrix of $ \cJ^3(Y_4) $ for a generic hypersurface is a matrix with the period matrices of the $ \cJ^1(R_\alpha) $ on the diagonal. These period matrices are independent due to the direct sum in \eqref{one_to_three-form}. At special points in complex structure moduli space, the lattice $ \Lambda $ of the intermediate Jacobian $ \cJ^3(Y_4) $ will degenerate
and require an extension of this diagonal ansatz. While we will not consider such phenomena in this work, it would be 
interesting to explore them in the future. The intermediate Jacobian admits a 
positive definite quadratic form $ Q $  introduced in \eqref{def-Q}. Evaluated for two $(2,1)$-forms $\psi_\cA$ and $\psi_\cB$, we recall that  
\beq \label{Q-expan}
   Q(\psi_\cA,\psi_\cB)  = - i v^\Sigma \int_{Y_4} \omega_\Sigma \wedge \psi_\cA \wedge \bar \psi_\cB \ ,
\eeq
where we inserted the expansion of $J= v^\Sigma \omega_\Sigma$ given in \eqref{exp-J}.
Note that we can pick a basis $\omega_\Sigma$ that is Poincar\'e-dual to a set of $h^{1,1}(Y_4)$ homologically 
independent divisors $D_\Sigma'$ of $Y_4$.\footnote{From the description of the Gysin map $\iota_{\Sigma \ast}$ given above, it is clear that  the $\omega_\Sigma$ can be written as $\omega_\Sigma = \iota_{\Sigma \ast}1$, $ 1 \in H^0(D^\prime _\Sigma, \mathbb{C}) $, for the embedding $\iota_{\Sigma} : D_\Sigma' \hookrightarrow Y_4$.}  
We will now evaluate the quadratic form $Q$ for the $(2,1)$-forms constructed in \eqref{chi_cA_Gysin}.

\begin{figure}
\begin{center}
\setlength{\unitlength}{0.7cm}
\begin{picture}(12,12)
\put(-1,0){\includegraphics[height=8.4cm]{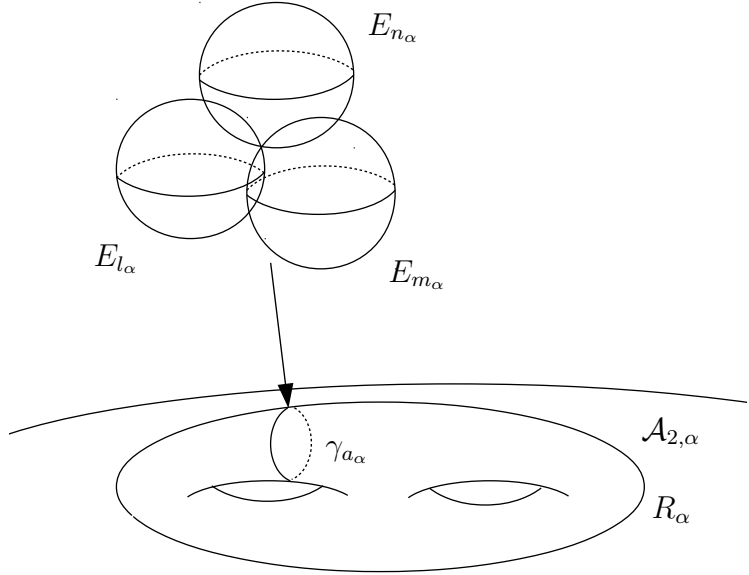} }
\put(11.2,1.5){$ R_\alpha $}
\put(11,3.0){$ \cA_{2,\alpha} $}
\put(5.0,2.7){$ \gamma_{a_\alpha} $}
\put(0.6,6.3){$ E_{l_\alpha} $}
\put(6.2,6.0){$ E_{m_\alpha} $}
\put(5.8,10.7){$ E_{n_\alpha} $}
\end{picture}
\caption{\textit{Intersection structure of the divisors $ D^\prime _{l_\alpha} $ that are fibration over $ R_\alpha $ with fiber $ E_{l_\alpha} $ and holomorphic one-forms $ \gamma_{a_\alpha} $.}}
\label{intersection_structure_picture}
\end{center}
\end{figure}

In order to do that, we first analyze the appearing intersection structures. Using \eqref{chi_cA_Gysin} we have associated the divisors $D'_\Sigma$, $D'_{l_\alpha}$, and $D'_{l_\beta}$ to 
the forms $\omega_\Sigma$, $\psi_\cA$, and $\psi_\cB$, respectively. 
We now claim that the integral in \eqref{Q-expan} is only non-zero if the curve 
\beq \label{def-cC}
     \cC = D'_\Sigma \cap D'_{l_\alpha} \cap D'_{l_\beta} 
\eeq
is in the same homology class as one of the Riemann surfaces $R_\alpha$ or $R_\beta$. In fact, we argue 
that all three divisors in \eqref{def-cC} have to be resolution divisors $D_{l_\alpha}'$ for the \textit{same}  
Riemann surface $R_\alpha$, i.e.~the only relevant intersections are
\beq \label{introduce_hatM}
   D'_{l_\alpha} \cap D'_{m_\alpha} \cap D'_{n_\alpha}= \hat M_{l_\alpha m_\alpha n_\alpha} \cdot R_\alpha \, ,
\eeq
where $ \hat M_{l_\alpha m_\alpha n_\alpha} $ are intersection numbers we discuss next.
 To see this we note that the intersection curve $\cC$ is again a hypersurface in the toric variety $ D_\Sigma \cap D_{l_\alpha} \cap D_{l_\beta} $. In order that it has non-trivial one-forms that lift to $Y_4$, it has to be two-semiample and hence corresponds to one of the Riemann surfaces $ R_\alpha $. Since all three divisors in \eqref{introduce_hatM} are fibrations of $E_{l_\alpha}$ over $R_\alpha$ we 
 can read off
 \beq
    E_{l_\alpha} \cap E_{m_\alpha} \cap E_{n_\alpha}= \hat M_{l_\alpha m_\alpha n_\alpha} \, .
 \eeq
 We depicted the intersection structure in figure \ref{intersection_structure_picture}.

Note that due to the fact that the $E_{l^\alpha}$ are realized as toric 
subspaces of  $\cA_5$ as noted around \eqref{Elalph_fibrD} and  our assumption that $ \cA_5 $ is smooth,  the intersection numbers $ \hat M_{l_\alpha m_\alpha n_\alpha}$ can be computed directly in $\cA_5$. This implies that they are
are either one or zero, i.e.~are the normalized volume of the face spanned by $ \nu^\ast _{l_\alpha} , \nu^\ast _{m_\beta} , \nu^\ast_{n_\alpha} $.
 Returning to the expansion of $Q$ in \eqref{Q-expan} it is convenient to chose an adopted 
 basis for the $J$ expansion 
\begin{equation}
 J = v^\Sigma \, \omega_\Sigma = \sum_{\alpha=1}^{n_2} \sum_{l_\alpha} v^{l_\alpha} [D^\prime _{l_\alpha}] + \ldots \, ,
\end{equation}
where we only displayed the $ v^{\Sigma} $ that will contribute to $ Q $. 
Putting everything together we then arrive at 
\begin{equation} \label{Q-hypersurfaces}
 Q(\psi_\cA, \psi_\cB) = - i \delta_{\alpha \beta}\ v^{l_\alpha} \hat M_{l_\alpha m_\alpha n_\beta} \int_{R_\alpha} \gamma_{a_\alpha} \wedge \bar \gamma_{b_\beta} \, ,
\end{equation}
for multi-indices $\cA = (\alpha, m_\alpha,a_\alpha)$ and $\cB = (\beta,n_\beta,b_\beta)$.
Another way to interpret this geometrically is to say that for a fixed $R_\alpha$ the corresponding $ E_l $-fibers 
form an analogue of the Hirzbruch-Jung sphere-tree, familiar from the resolution of codimension two orbifold 
singularities, and the precise intersection pattern $\hat M_{l m n}$ of these fibers determines the bilinear 
form $ Q $. Hence, $ Q $ depends on the triangulation of the ambient space $ \cA_5 $. The dependence on K\"ahler moduli is contained in the structure of this higher dimensional sphere-tree. The complex structure dependence of $ Q $ can be fully reduced to the complex structure dependence of the one-forms on the Riemann surfaces $ R_\alpha$.

Having evaluated the quadratic form $Q$ for the geometries under consideration, it is now straightforward to 
read off the holomorphic function $f_{\cA \cB}$ and the constants ${M_{\Sigma \cA}}^\cB$, ${M_{\Sigma }}^{\cA \cB}$ defined in \eqref{def-MM}. Comparing the general expression \eqref{Qpsipsi} to our result \eqref{Q-hypersurfaces} we first realize that 
\beq \label{M_vanish}
       {M_{\Sigma }}^{\cA \cB} = 0 \, .
\eeq
To see this we denote by $\hat f^{(\alpha)}_{a_\alpha b_\alpha}$ the holomorphic function associated to $R_\alpha$. The 
equation \eqref{eval_gammagamma} then reads
\begin{equation} \label{gammagamma_alpha}
- i\int_{R_\alpha} \gamma_{a_\alpha} \wedge \bar \gamma_{b_\alpha} = 2 \cdot \text{Re}\, \hat f^{(\alpha)}_{a_\alpha b_\alpha}\ .
\end{equation}
In contrast to \eqref{Qpsipsi} only the real part of $\hat f^{(\alpha)}_{a_\alpha b_\alpha}$ appears. In other 
words, the vanishing condition \eqref{M_vanish} arises from the fact that one can chose a canonical basis $(\hat \alpha_{a_\alpha},\hat \beta^{a_\alpha})$, 
as defined in \eqref{sympl_on_R}, on each $R_\alpha$. To read off $f_{\cA \cB}$ and ${M_{\Sigma \cA}}^\cB$ one has the 
freedom of multiplying with a constant matrix, which corresponds to choosing a different basis $(\alpha_\cA,\beta^\cA)$
in \eqref{exp-psicA}. 
A convenient way to chose a basis is to use the pullback and Gysin maps as in \eqref{chi_cA_Gysin}, i.e.~we define
\beq \label{lift-hatalpha}
   \alpha_\cA = \iota_{l_\alpha \ast}(\pi^*_{l^\alpha} \hat \alpha_{a_\alpha})\ , \qquad 
   \beta^\cA = \iota_{l_\alpha \ast}(\pi^*_{l^\alpha} \hat \beta^{a_\alpha})\ , 
\eeq
with multi-index $\cA = (\alpha,l_\alpha,a_\alpha)$. The claim that all moduli dependence is captured by  
the periods of $R_\alpha$ is equivalent to the statement that the so-constructed $(\alpha_\cA,\beta^\cA)$ 
are independent of the moduli. This moduli-independence is a requirement in the general 
construction of \autoref{sec:threeform-general}. With \eqref{lift-hatalpha} one checks again \eqref{M_vanish} and 
computes 
\beq \label{McAcB}
    {M_{\Sigma \cA}}^\cB = \left\{ \begin{array}{ccc} \hat M_{l_\alpha m_\alpha n_\alpha} \delta_{a_\alpha}^{b_\alpha} & & \text{for} \, \, \alpha =\beta\ \text{and} \ \Sigma= l_\alpha \\[.1cm]
                                                                                      0 & & \text{otherwise}\, ,\end{array} \right. 
\eeq
with mulit-indices $\cA=(\alpha,m_\alpha,a_\alpha)$ and $\cB=(\beta,n_\beta,b_\beta)$.
Inserting this expression into \eqref{Qpsipsi} and comparing with \eqref{Q-hypersurfaces} using 
\eqref{gammagamma_alpha} we finally read off 
\beq \label{f_final}
     f_{\cA\cB} = \left\{ \begin{array}{ccc} \hat f_{a_\alpha b_\alpha}^{(\alpha)} \delta_{m_\alpha n_\alpha} & & \text{for} \, \, \alpha =\beta\ \\[.1cm]
                                                                                      0 & & \text{otherwise}\, , \end{array} \right.
\eeq
with mulit-indices $\cA=(\alpha,m_\alpha,a_\alpha)$ and $\cB=(\beta,n_\beta,b_\beta)$.

The identifications \eqref{M_vanish}, \eqref{McAcB}, and \eqref{f_final} together with the computations of $\hat f^{(\alpha)}_{a^\alpha b^\beta}$ in \autoref{Riemann_periods} constitute our main results for the 
analysis of Calabi-Yau fourfold hypersurfaces in toric varieties. We find that $f_{\cA \cB}$ actually factors into 
non-trivial blocks, each containing the information about one of the embedded Riemann surfaces. The non-trivial 
couplings $ {M_{\Sigma \cA}}^\cB $ capture the intersection information of the generalized sphere-tree over each Riemann surface. 
It is worthwhile to stress that this information suffices to compute the crucial parts of the effective actions 
relevant, for example, in \cite{Grimm:2014vva,Carta:2016ynn}. However, it is also clear that certain applications 
will require to consider a more general class of geometries. For example, the non-Abelian structures 
considered in \cite{Grimm:2015ona,Grimm:2015ska,Corvilain:2016kwe} are expected to require 
the use of complete intersections and to find less block-diagonal situations. 
We hope to return to such more involved geometries in the future.

\subsection{Three-form periods on Fermat hypersurfaces in weighted projective spaces}  \label{subsec:weighted_periods}

To close our discussion on the construction of three-form periods on Calabi-Yau fourfolds, we examine 
a particularly simple class of geometries, Fermat hypersurfaces in weighted projective spaces. 
Since weighted projective spaces are the simplest examples of toric varieties, the concepts introduced in the previous section
apply directly and can be more intuitively understood. The explicit examples investigated in \autoref{sec:Example} will 
also fit into this class of geometries.

Since we are primarily interested in the calculation of the normalized period matrix $ f_{\cA \cB} $, which was  shown in \eqref{f_final} to only depend on the chiral ring of a Riemann surface embedded in the hypersurface, it is not necessary to blow-up orbifold singularities.  Therefore, our analysis of the geometries simplifies drastically. The exact pattern of blow-ups necessary to produce a smooth ambient space only enters through the intersection numbers $ {M_{\Sigma \cA}}^\cB $ determined in \eqref{McAcB}. This will enable us to discuss the derivation of the Picard-Fuchs equation explained in \autoref{Riemann_periods} more explicitly. Afterwards, we will discuss the case when all three-forms are induced by a single divisor.

In the following we focus on a generally singular ambient space $\cA_5$, which is a weighted projective spaces $ \cA_5 = \mathbb{P}^5(w_1, \ldots,w_5, w_6 = 1) $ realized by a simplicial polyhedron in $ N_{\mathbb{Q}} = \mathbb{Q}^5 $ with six vertices
\begin{equation}
 \nu^\ast _i = e_i, \, , \quad i = 1, \ldots, 5 \, ,  \quad \nu^\ast _6 = (-w_1,-w_2,-w_3,-w_4,-w_5) \, .
\end{equation}
The choice of $ w_6 = 1 $ enables us to express all toric divisors $ [D _i] $ as a multiple of $ [D_6] = [H] $,
\begin{equation}
 [D_i] = w_i [H] \, ,
\end{equation}
that can be viewed as a generalization of the hyperplane class one encounters in classical projective spaces. In the homogeneous coordinate ring
\begin{equation}
 S = \mathbb{C}[X_1, \ldots, X_6] \, ,
\end{equation}
we hence obtain the usual grading of a monomial by a positive number, the multiple of $ H $ it corresponds to.

The anti-canonical hypersurface $ Y^{\rm sing}_4 $ in $ \cA_5 $ is given by the zero set of a degree $ d $ polynomial, with $ d $ such that
\begin{equation}
 w_i | d, \ i = 1, \ldots, 6 \, , \qquad \sum D_i = - K_{\cA_5} = d \cdot H \, .
\end{equation}
The first condition allows the hypersurface to be a deformation of a Fermat hypersurface. In particular, it enables us to choose the non-degenerate hypersurface in the equivalence class of the anti-canonical divisor to be
\begin{equation}
 p_\Delta =  X_1 ^{d/w_1} + \ldots + X_6 ^{d/w_6} + \sum_{\nu \in \Delta, \, \text{codim}(\nu)>1} a_\nu \, p_\nu \, .
\end{equation}
where the six vertices spanning the polyhedron $ \Delta \subset \mathbb{Q}^5 $ are given by
\begin{equation}
 \nu_i = - \sum e_j + \frac{d}{w_i} e_i \in \mathbb{Z}^5 \, , \quad i = 1, \ldots , 5 \, , \quad \nu_6 = - \sum e_j \in \mathbb{Z}^5 \, .
\end{equation}
Due to the assumption of a the existence of a Fermat surface in the equivalence class of the anti-canonical divisor, $ \Delta $ is a simplex. This is not true for a general toric ambient space and is a rather restrictive assumption.

In this situation, a surface $ \cA_2 $ of $ \mathbb{C}^3 / \mathbb{Z}_n $-singularities in the ambient space arises if exactly three weights have a common divisor $ n $. Without loss of generality we can assume
\begin{equation}
 n \,|\, w_3, w_4, w_5 \, , \quad n \, \not| \, \, w_1, w_2, w_6 \, ,
\end{equation}
i.e. $ \cA_2 $ is given as the subspace of $ \cA_5 $ given by $ X_1 = X_2 = X_6 = 0 $. This $ \mathbb{C}^3 / \mathbb{Z}_n $-singularity will lead to a curve $ R $ of $ \mathbb{C}^3 / \mathbb{Z}_n $-singularities in the hypersurface $Y_4^{\rm sing}$ that intersects $ \cA_2 $ transversely and clearly requires a number of blow-ups to resolve this singularity. The corresponding divisors will induce the non-trivial three-forms on the smooth hypersurface $Y_4$, but its complex structure dependence will be fully captured by the curve $ R $ of $ \mathbb{C}^3 / \mathbb{Z}_n $-singularities.

Our ansatz implies in particular that $ n | d $. The toric surface $ \cA_2 $ of $ \mathbb{C}^3 / \mathbb{Z}_n $-singularities is also a weighted projective space
\begin{equation}
 \cA_2 = \mathbb{P}^2(w_3, w_4, w_5) \simeq \mathbb{P}^2(w_3 /n, w_4/n, w_5/n) \, .
\end{equation}
This identification can be seen from the fact that the weights of $ \cA_2 $ are all multiples of $ n $ and only the ratio of two weights in a weighted projective space matters.
The corresponding hypersurface is just the restriction of the polynomial to this space, i.e.~setting $ X_1 =  X_2 = X_6 = 0 $ and hence $ R $ is isomorphic to
\begin{equation}
 R = \mathbb{P}^2(w_3/n, w_4/n, w_5/n) [d/n] \, ,
\end{equation}
i.e.~a degree $ d/n $-dimensional Fermat hypersurface in $ \cA_2 $.
In terms of lattice-polytopes, we find that the dual polyhedron of $ \cA_2 $ defined by $ \theta $ is in general not reflexive, it contains $ \ell^\prime(\theta) \geq 0 $ interior points and the genus of $ R $ is exactly the number of these interior points $ \ell^\prime (\theta) = g $. The Fermat polynomial on $ \cA_2 $ is given by the corresponding restriction of $ p_\Delta $ and reads
\begin{align}
 p_\theta &= \sum_{E_i \cap \theta^\ast} X_i ^{d/w_i} + \sum_{\nu_b \in \text{int}(\theta)} a_b p_b\nonumber \\
	  &= X_3 ^{d/w_3} + X_4 ^{d/w_4} + X_5 ^{d/w_5} + X_3 X_4 X_5 \Big( \!\!\! \! \sum_{\tiny \begin{array}{c}{{\text{deg}(p^\prime _b) =} }\\ w_1 + w_2 + w_6\end{array}}\!\!\!\! a_b p^\prime_b(X_3, X_4, X_5) \Big) \, .
\end{align}
where we introduced the monomials
\beq
p^\prime _a \in \cR_\theta (K_{\cA_5} |_{\cA_2} - K_{\cA_2}) = \cR_\theta (w_1 + w_2 + w_6) \, ,
\eeq
which are the non-trivial monomials of $ \cR_\theta $ of degree $ w_1 + w_2 + w_6 $ corresponding to the integral interior points of $ \theta $, $ \nu_a \in \text{int}(\theta) \cap N $.

Following the construction of holomorphic one-forms on $ R $ outlined in \autoref{Riemann_periods}, we already seen how to construct $ \cR_{\theta} $ and we are left with the construction of the holomorphic volume-form of $ \cA_2 $.
The holomorphic volume-form $ d\omega_{\cA_2} $ of \eqref{holomorphic_two_form} is given by
\begin{equation} \label{domegaA2_W}
 d\omega_{\cA_2} = w_3 X_3 dX_4 \wedge dX_5 - w_4 X_4 dX_3 \wedge dX_5 + w_5 X_5 dX_3 \wedge dX_4 \, \in \Omega^2 _{\cA_2} ,
\end{equation}
and has degree $ w_3 + w_4 + w_5 $. The construction ensures the meromorphic two-forms
\begin{equation}
 \frac{p^\prime _a}{p_{\theta}} \, d\omega_{\cA_2} \, \in H^0 (\cA_2, \Omega^2 (R)) \, ,
\end{equation}
are globally defined on $ \cA_2 $. This means that they are invariant under the quasi-projective equivalence of the weighted projective space, i.e.~they have degree zero. In addition they have a first order pole along the Riemann surface $ R $, which facilitates the residue construction we introduced.

Therefore, we extracted all quantities needed to define the $ (1,0) $-forms $ \gamma_a $ of our ansatz
\begin{equation} \label{one_form_wproj}
 \gamma_a = \int_{\Gamma} \frac{p^\prime _a}{p_{\theta}} \, d\omega_{\cA_2} \, , \quad \nu_a \in \text{int}(\theta) \cap M\, .
\end{equation}
The next step to find the Picard-Fuchs equations, is to imply the relations in $ \cR_\theta $ to reduce the second derivatives of $ \gamma_a $ with respect to the complex structure moduli $ a_b $. In practice, however, this is connected with a significant amount of work, the number of relations goes with $ g^2 $, which should be attempted via an adapted algorithm that suits an implementation in a computer program. We will outline the calculation for the simplest example, $ g=1 $, in the upcoming section.

For generic orbifold singularities along a curve $ R $ in a toric Calabi-Yau fourfold hypersurface $ Y_4 $, we encounter in general complicated intersection patterns of the necessary toric blow-ups, which we however need to understand to calculate the intersection numbers $ {M_{\Sigma \cA}}^\cB $, \eqref{McAcB}.

\begin{figure}
\begin{center}
\setlength{\unitlength}{0.7cm}
\begin{picture}(12,12)
\put(0,0){\includegraphics[height=8.4cm]{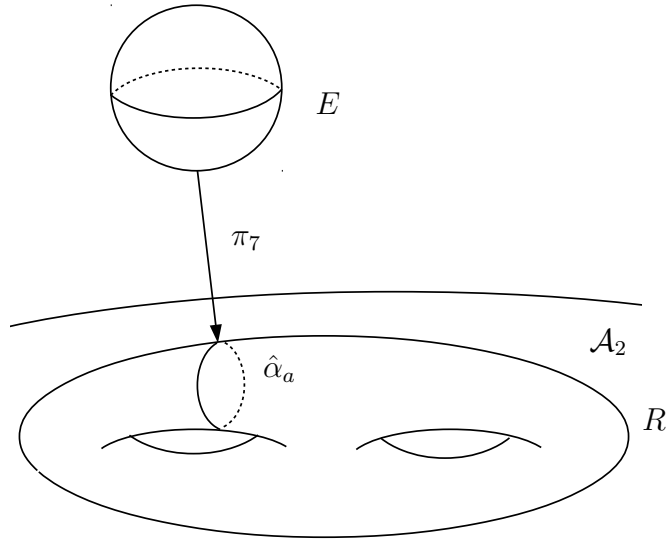} }
\put(12.0,3){$ R $}
\put(11,4.5){$ \cA_2 $}
\put(4.2,6.5){$\pi_{7} $}
\put(4.8,4.0){$\hat \alpha_a $}
\put(5.8,9.0){$ E $}
\end{picture}
\caption{\textit{Fibration structure of $ D^\prime _7 $. The Riemann surface $ R $ is a hypersurface of the toric space $ \cA_2 $ over which the toric surface $ E $ is fibered.}}
\label{fibration_structure_picture}
\end{center}
\end{figure}

The simplest case of an orbifold singularity $ \mathbb{C}^3 / \mathbb{Z}_n $ along the Riemann surface $ R $ is a $ \mathbb{C}^3 / \mathbb{Z}_3 $-singularity, i.e.~$ n=3 $, that can be resolved by one toric blow-up and as a result we obtain a divisor $ D^\prime _7 = \{ X_7 = 0 \} $ that is a fibration over the Riemann surface $ R $ with exceptional fiber $ E $. The corresponding additional ray $ \tau_7 $ goes through the integral point $ \nu^\ast _7 $
\begin{equation}
 \nu^\ast _7 = \frac{1}{3}(\nu^\ast _1 + \nu^\ast _2 + \nu^\ast _6)
\end{equation}
and the fiber $ E $ is just
\begin{equation}
 E = \mathbb{P}^2(w_1,w_2,w_6) \, ,
\end{equation}
which is for general $ w_1, w_2 $ not smooth. Resolving the corresponding point singularity leads to non-trivial three-cycles on $ D^\prime _7 $ that will be trivial in $ Y_4 $. Since we have only one blow-up divisor $ D^\prime _7 $ resolving the $ \mathbb{C}^3 / \mathbb{Z}_3 $-singularity along the curve $ R $, the intersection matrix $ {M_{\Sigma \cA}}^\cB $, \eqref{McAcB}, simplifies drastically, to the single number $ M = 1 $. In this situation, all non-trivial three-forms $ \psi_\cA $ of the smooth hypersurface $ Y_4 $ arise from $ D^\prime _7 $ and correspond to a one-form $ \gamma_a $ on $ R $. The multi-index $ \cA $ runs only over the one-forms on $ R $, $ \gamma_a \in H^0(R,\Omega^1)$, i.e.~$ \cA = (\alpha, l_\alpha, a_\alpha) = (1,7,1), \ldots , (1,7,g) $. Correspondingly, we find using \eqref{chi_cA_Gysin}
\begin{equation}
 \psi_\cA =  \iota_{7 \ast} \big( \pi_7 ^\ast \, \gamma_{a} \big) \, \in H^{2,1}(Y_4) \, ,
\end{equation}
and hence for the positive bilinear form $ Q $, \eqref{Q-hypersurfaces}, that
\begin{equation}
 Q(\psi_\cA, \bar \psi_\cB) = 2\,  v^7 \cdot \R \hat f_{ab} \, ,
\end{equation}
with $ \hat f_{ab} $ the normalized period matrix of $ R $, that can be calculated via the Picard-Fuchs equations, and $ v^7 $ is the volume modulus associated to the Poincar\'e dual two-form of $ D_7 ^\prime $. We end this discussion with a schematic sketch of the fibration structure of $ D^\prime _7 $ we encountered in this example, \autoref{fibration_structure_picture}. The reader should keep this picture in mind, when we discuss explicit geometries in the next section.

\section{Calabi-Yau hypersurface examples} \label{sec:Example}

In this section we discuss two simple Calabi-Yau fourfold examples with non-trivial three-form cohomology. 
In the course of this analysis we will encounter several consequences of these non-trivial three-forms 
when using the geometry as F-theory background. In particular, we will investigate the weak-coupling limit of 
Sen and trace some of the properties of the three-form moduli and their couplings through this limit. 
Our findings provide further motivation to explore regions in complex structure moduli space that do not 
yield weakly coupled Type IIB orientifold backrounds.
 
\subsection{Generalities}

To begin with, we will discuss general aspects of the effects of non-trivial three-form cohomology in F-theory.
We keep our considerations simple, by focusing on elliptically fibered Calabi-Yau fourfolds realized as hypersurfaces in weighted projective spaces as discussed in \autoref{subsec:weighted_periods}. For general hypersurfaces the three-form moduli $\cN_\cA$ yield 
complex scalar fields in the four-dimensional effective theory. These scalars can have two interrelated origins in a general F-theory setting: (1) they can arise as zero-modes of the R-R and NS-NS two-forms, or (2) they can correspond to continuous Wilson line moduli arising on seven-branes. In general this distinction can be meaningless (see e.g.~\cite{Douglas:2014ywa}), 
but it becomes more stringent in the weak string coupling limit. 
We will encounter both types of moduli in two simple example geometries in \autoref{two-form-scalars} and \autoref{Wilson_scalars}.

\subsubsection{Weierstrass-form and non-trivial three-form cohomology}

Let us consider a Calabi-Yau hypersurface $ Y_4^{\rm sing}$ in a weighted projective space $ \cA_5 = \mathbb{P}^5(w_1,\ldots,w_6=1) $. As we have seen in \autoref{subsec:weighted_periods}, we can find after a resolution of $Y_4$ in $\hat \cA_5$ a 
smooth Calabi-Yau manifold with non-trivial three-form cohomology. We have discussed in detail that the 
complex structure dependence of these three-forms can already be inferred from the complex structure variations of one-forms on Riemann surfaces embedded in $ Y_4^{\rm sing}$.

To obtain an F-theory background we want to consider an elliptically fibered Calabi-Yau fourfold with a section. Therefore, 
we specialize to Weierstrass-models with the elliptic fiber realized as hypersurface in 
$ \cA_{fiber} = \mathbb{P}^2(1,2,3) $ fibered over a toric basis $ B_3 $ that will be a (blow-up of a) weighted projective space
\begin{equation}
 B_3 ^{\rm sing} = \mathbb{P}^3(w_1,w_2,w_3,w_6=1) \, .
\end{equation}
The blow-up may be necessary for example to obtain generalized Hirzebruch surfaces, i.e.~$ \mathbb{P}^1 $-fibrations over a two-dimensional toric variety. Note here that since we assume the base $ B_3 $ to be toric, in can not carry non-trivial three-form cohomology. In general the polyhedron of the base $ \Delta^\ast _{base} $ is not convex. This implies that 
the base is non-Fano and a resolution of singularities might involve choices of an extension of $ \Delta^\ast $ by integral 
vertices in its interior or exterior. Clearly this will alter the geometry of the corresponding Calabi-Yau fourfold and 
also the resulting physics.

The elliptically fibered Calabi-Yau fourfold $ Y_4 ^{\rm sing}$ is 
determined by the data of the base $ B_3 ^{\rm sing}$ and is a hypersurface in
\begin{equation}
 \cA_5= \mathbb{P}^5(w_1,w_2,w_3, w_4 = 2w , w_5 = 3w , w_6=1)\ , \quad w = w_1 + w_2 + w_3 + w_6 \, .
\end{equation}
It is common to denote the projective coordinates as $ X_4 = x $ and $ X_5 = y $. 
The vanishing of the first Chern-class of the hypersurface $ Y_4 ^{\rm sing}$ requires the defining polynomial to have Tate form with degree $ d = 6 \, w $ given by 
\begin{equation} \label{long_weierstrass}
 y^2 + a_1 xy + a_3 y = x^3 + a_2 x^2 + a_4 x + a_6\ .
\end{equation}
Here $ a_j $ are global sections of various powers of the anti-canonical bundle $ K_{B_3} ^{-1} $ of the base $ B_3 $:
\begin{equation}
 a_j \in H_0(B_3, K^{-j} _{B_3}) \, .
\end{equation}
On the singular space this simply requires the $ a_j $ to be quasi-homogeneous polynomials in the projective base coordinates of degree $ \text{deg}(a_j) = w \cdot j $. After performing the blow-ups the structure of these global sections $ a_j $ may be more complicated since toric blow-ups changing $ \Delta^\ast $ will in general also affect $ \Delta $ and hence the anti-canonical divisor class providing the Calabi-Yau hypersurface.

\subsubsection{The weak string coupling limit} \label{sen_limit}

Let us next recall the weak string coupling limit in complex structure moduli space following Sen \cite{Sen:1997bp,Sen:1997gv} and \cite{Clingher:2012rg} for a more refined version. By a variable redefinition, we can bring any Tate form \eqref{long_weierstrass} into standard Weierstrass form given by 
\begin{equation}
 y^2 = x^3 + fx + g \, .
\end{equation}
In order to do that we note that $ f \in \, H^0(B_3,K_{B_3} ^{-4}) $  $\in \, H^0(B_3,K_{B_3} ^{-6})$ 
and can be written as
\beq \label{fg-bdecomp}
 f = - \frac{1}{48}(b_2 ^2 - 24 \epsilon \, b_4)  \ , \qquad
 g = - \frac{1}{864}(- b_2 ^3 + 36 \epsilon \, b_2 b_4 - 216  \epsilon^2 \, b_6)  \ ,
\eeq
with $ b_i $ global sections of $ K_{B_3} ^{-i} $. In our conventions the $ b_i $ are related to the $ a_j $ of the Tate form \eqref{long_weierstrass} via
\beq \label{bi-def}
 b_2 = a_1 ^2 + 4 a_2 \ , \qquad 
 b_4 = a_1 a_3 + 2 a_4 \ , \qquad
 b_6 = a_3 ^2 + 4a_6\ .
\eeq
The parameter $ \epsilon $ introduced in \eqref{fg-bdecomp} can be thought of as the complex structure modulus 
that needs to be sent to zero to perform the weak string coupling limit. We discuss this limit in more detail next. 

If one starts from an F-theory compactification on the smooth elliptically fibered $ Y_4 $ there is a corresponding weak string coupling configuration that only admits D7-branes and O7-planes. 
This weak coupling limit is obtained by sending $ \epsilon \rightarrow 0 $. To see this one notes that the complex structure $\tau$ of the elliptic fiber is given by 
\beq \label{j-Delta}
   j(\tau) =  \frac{4 (24 f)^3}{\Delta} \ , \qquad \Delta = 27 g^2 +4 f^3\ , 
\eeq
where $\Delta$ is the discriminant dictating the locations in the base along which the fiber degenerates. Inserting \eqref{fg-bdecomp} into \eqref{j-Delta} one expands
\beq \label{Delta-exp}
  \Delta = \frac{1}{64} \epsilon^2 b_2^2 (b_2 b_6 - b_4 ^2) \ , \qquad j(\tau)= -\frac{32 b_2^4}{(b_2 b_6-b_4^2 ) \epsilon^2}\ ,
\eeq
where we only displayed the leading terms. This implies that in the limit $ \epsilon \rightarrow 0 $ $ \I \, \tau \propto - \log \epsilon $ everywhere expect at the locus $b_2 = 0$. Recalling that in Type IIB supergravity one has $\tau = C_0 + i e^{-\phi}$, with $e^{\langle \phi \rangle} = g_s$, we thus conclude that $g_s \rightarrow 0$ in the limit $\epsilon \rightarrow 0$.

The extended objects in the weak coupling configurations are D7-branes and O7-planes. Using the split \eqref{Delta-exp} of the discriminant one identifies the following locations of the $ D7 $-branes and $ O7 $-planes:
\beq
\text{O7} : \quad b_2 = 0 \, , \qquad \text{D7}: \quad b_2 b_6 - b_4 ^2 = 0 \, .
\eeq 
The corresponding Calabi-Yau threefold $ Y_3  $ is a double cover of the toric base $ B_3 $ with branching locus the O7-planes. In practice $ Y_3 $ is obtained as a hypersurface in the anti-canonical line-bundle $ K_{B_3} ^{-1} $ with fiber coordinate $ \xi $ and equation
\beq
Y_3: \quad Q = \xi^2 - b_2 = 0 \, .
\eeq 
The orientifold involution acts as $\sigma: \, \xi \rightarrow -\xi$ in this equation, such that $\xi = b_2 = 0$ is indeed the fixed-point set determining the location of the O7-planes. 

For $ B_3 $ a blow-up of a weighted projective space $ B^{\rm sing} _3 = \mathbb{P}(w_1,w_2,w_3,w_6=1) $ this implies that $ Y_3 $ can be embedded in the corresponding blow-up $ \hat \cA_4 $ of
\begin{equation}
 \cA_4  = \mathbb{P}(w_1,w_2,w_3,w_6=1,w) \, , \quad w = w_1 + w_2 + w_3 + 1 \, ,
\end{equation}
as an anti-canonical hypersurface of degree $ 2w $. The (resolved) toric ambient space $ \hat \cA_4 $ is a $ \mathbb{P}^1 $-fibration over $ B_3 $. Therefore, we can apply  toric geometry techniques to analyze this setting.

Let us also discuss the five-cycles in $ Y_4 $ that lead by Poincar\'e duality to non-trivial three-forms. Due to the fact that the base $ B_3 $ of our elliptic fibration is toric, all three-forms need to have one leg in the fiber, i.e. the dual five-cycles are circle fibrations with the circle a cycle of the elliptic fiber. As discussed in \cite{Denef:2008wq} expanding the three-form potential of M-theory leads in this case to
\begin{equation} \label{C_3expansion}
 C_3 = B_2 \wedge dx + C_2 \wedge dy + \ldots
\end{equation}
where $ dx $ and $ dy $ are a basis of one-forms on the elliptic fiber dual to its two one-cycles $ A,B $. Comparing with the discussion in \autoref{sec:three-forms_hyper} this implies that for the very special case that $ R_\alpha $ is the elliptic fiber, the divisors $ D^\prime _\alpha $ are all direct products $ D^\prime _\alpha = R_\alpha \times E_{l_\alpha} $ where $ E_{l_\alpha} $ is a base divisor. The period matrix of $ R_\alpha $ is hence $ i \tau $ the axio-dilaton that is constant over the divisor $ E_{l_\alpha} $. In the weak coupling limit we find that the three-form moduli can be identified with the so called odd moduli $ G^\cA = G^{l_\alpha} $ where
\begin{equation}
G = G^{l_{\alpha}} \omega_{l_{\alpha}} = B_2 + i \tau \, C_2 \, \in \, H^{1,1} _- (Y_3) \, .
\end{equation}
For details on the odd moduli of a Calabi-Yau threefold $ Y_3 $ we refer to \cite{Grimm:2005fa}.

There is, however, a second kind of five-cycle. These are circle-fibrations over four-chains in the base that degenerate at the boundaries of the chain. Think of these cycles in the way a two-sphere is a circle fibration over an interval that degenerates at the two endpoints of the interval. In physical language the four-chains have their boundaries on seven-branes, where the elliptic fibration degenerates, that wrap divisor of the base $ B_3 $. On the seven-branes we hence find three-cycles that are dual to one-forms on the divisors. Due to this geometrical picture and including the monodromy properties of the resulting cohomology classes we hence infer that the three-forms we constructed split in the weak coupling limit into two classes
\beq
    H^{2,1}(Y_4) \quad \longrightarrow \quad  \left\{ \begin{array}{c} H^{1,1} _- (Y_3) \, , \\[.1cm]
                                                                                      H^{1,0} _- (S) \, .\end{array} \right. 
\eeq
Here $ S $ is a divisor in $ Y_3 $ wrapped by a D7-brane. The monodromy properties can be deduced from the fact that both one-cycles of the elliptic fiber are odd under the orientifold involution. An example for the first case will be discussed in \autoref{two-form-scalars} while we present in \autoref{Wilson_scalars} an example for the second case.

\subsection{Example 1: An F-theory model with two-form scalars} \label{two-form-scalars}

In this subsection we introduce the first example geometry. It admits only one cohomologically 
non-trivial $(2,1)$-form such that its moduli dependence can be described by a two-torus. It turns 
out that this two-torus is actually the elliptic fiber over a specific divisor in the base. We will thus be able 
to discuss the three-form periods and weak string coupling limit in detail. 

\subsubsection{Toric data and origin of non-trivial three-forms} \label{toric_data_example1}

The first example of an elliptically fibered Calabi-Yau fourfold with non-trivial three-forms 
appeared already in the list of hypersurfaces in weighted projective spaces in \cite{Klemm:1996ts}.
It is constructed by starting with the weighted projective space $ \cA _5 = \mathbb{P}^5[1,1,1,3,12,18]$, 
which is singular due to the fact that the last three weights have a common divisor $3$ and the last two have a common divisor $2$.
The former property yields $\bbC^3 / \bbZ_3$-singularities along a surface $\cA_2$ in $\cA_5 $, while the latter results in $ \bbC^4/\bbZ_2$-singularities along a curve in $\cA_5 $.
The anti-canonical hypersurface $ Y_4 ^{\rm sing} $  in $ \cA_5 $ is given by a 
polynomial $ p_\Delta $ of quasi-homogeneous degree $ 36 $.
Let us introduce complex projective coordinates on $\cA_5$ as $[\underline{u}: w:x:y]$ with the abbreviation $\underline{u}=(u_1,u_2,u_3)$.
The most general hypersurface equation of this type always can be brought to the form 
\beq
 p ^{\rm sing} _\Delta = y^2 + x^3 + \hat a_1\, x y + \hat a_2 \, x^2 +\hat a_3 \, y + \hat a_4 \, x  + \hat a_6 = 0\ ,
\eeq
with 
\beq \label{hatan}
 \hat a_n  = \sum^{2n}_{m=0} w^{2n-m}\, c_{n,m}(\underline{u})\ ,
\eeq
where $c_{n,m}(\underline{u})$ are general homogenous polynomials of  degree $3m$ in $\underline{u}=(u_1,u_2,u_3)$. 
Note that setting $u_1=u_2=u_3=0$
one finds the curve
\beq
   y^2+ x^3 + \hat c_1\, x y + \hat c_2 \, w^4 x^2 +\hat c_3 \, w^6 y + \hat c_4 \, w^8 x  + \hat c_6  w^{12} =0\ , 
\eeq
where $ \hat c_n = c_{n,0}$ are constants. Along this curve we have $ \bbC^3/\mathbb{Z}_3 $-singularities in the hypersurface $ Y_4 ^{\rm sing}$ of $ \cA_5 $.
 
We can resolve the $\bbZ_2,\bbZ_3$ singularities of the ambient-space $\cA_5$ by moving to a toric space $\hat \cA_5$. The fan of $ \hat \cA_5$ is (uniquely) determined by the cones with rays
\begin{equation} \label{poly_1}
\begin{tabular}[0]{|c  r  r  r  r  r  r  | c | c | c | c|}\hline
   \multicolumn{7}{|c|}{\text{Example 1: \ \ Toric data of $ \hat \cA_5 $}}    &\rule[-.2cm]{0cm}{.7cm} coords & $ \ell_1 $ & $ \ell_2 $ & $ \ell_3 $ \\
  \hline
    $ \nu^\ast_1 = $ & $ ( $ & $ 1$ & $ 0 $ & $ 0 $ & $ 0 $ & $ 0 ) $ & $ z_1 = u_1 $ & $ 0 $ & $ 1 $ & $ 0 $\\	
    $ \nu^\ast_2 = $ & $ ( $ & $ 0$ & $ 1 $ & $ 0 $ & $ 0 $ & $ 0 ) $ & $ z_2 = u_2 $ & $ 0 $ & $ 1 $ & $ 0 $\\
    $ \nu^\ast_3 = $ & $ ( $ & $ 0$ & $ 0 $ & $ 1 $ & $ 0 $ & $ 0 ) $ & $ z_3 = w   $ & $ 0 $ & $ 0 $ & $ 1 $\\
    $ \nu^\ast_4 = $ & $ ( $ & $ 0$ & $ 0 $ & $ 0 $ & $ 1 $ & $ 0 ) $ & $ z_4 = x   $ & $ 2 $ & $ 0 $ & $ 0 $\\
    $ \nu^\ast_5 = $ & $ ( $ & $ 0$ & $ 0 $ & $ 0 $ & $ 0 $ & $ 1 ) $ & $ z_5 = y   $ & $ 3 $ & $ 0 $ & $ 0 $\\
    $ \nu^\ast_6 = $ & $ ( $ & $-1$ & $-1 $ & $-3 $ & $-12$ & $-18) $ & $ z_6 = u_3 $ & $ 0 $ & $ 1 $ & $ 0 $\\	
    $ \nu^\ast_7 = $ & $ ( $ & $ 0$ & $ 0 $ & $-1 $ & $-4 $ & $-6 ) $ & $ z_7 = v   $ & $ 0 $ & $-3 $ & $ 1 $\\
    $ \nu^\ast_8 = $ & $ ( $ & $ 0$ & $ 0 $ & $ 0 $ & $-2 $ & $-3 ) $ & $ z_8 = z   $ & $ 1 $ & $ 0 $ & $-2 $\\
    \hline
\end{tabular} \, 
\, .
\end{equation}
Here we denoted by $ \ell_i $ the three projective relations between the coordinates, where we did not choose a minimal set of generators, like for the Mori-cone, but we have chosen a weight representation that emphasizes the fibration structure of the blown-up ambient space $ \hat \cA_5 $. It can be shown, as done in \cite{Batyrev:1994hm}, that this new ambient space only contains singular points and hence a general anti-canonical hypersurface is smooth.

The Calabi-Yau hypersurface $ Y_4$ is defined by a generic polynomial $ p_\Delta $ transforming as a section of the anti-canonical bundle $-K_{\hat \cA_5}$.
Translating the toric data \eqref{poly_1} into a hypersurface equation one finds that it takes the Tate form
\beq
 p_\Delta = y^2 + x^3 + a_1 x y z + a_2 x^2 z^2 + a_3 y z^3 +a_4 x z^4 + a_6 z^6=0\ ,
\eeq
where $[x:y:z]$ are the coordinates introduced in \eqref{poly_1} and the $a_i$ depend on the remaining coordinates. 
Hence, we infer that $ Y_4$ is an elliptic fibration over a toric base $ B_3$ with coordinates $[u_1:u_2:u_3:v:w]$ and elliptic fiber realized in $\bbP^2(2,3,1)$ with coordinates $[x:y:z]$.
Explicitly the $a_n$ are given by 
\beq \label{a_example_1}
   a_n = \sum_{m=0}^{2n} c_{n,m}(\underline{u})\  w^{2n - m} v^m \ ,
\eeq
where $c_{n,m}$ are homogeneous of degree $3m$ in the variables $\underline u=(u_1,u_2,u_3)$. It is instructive to point out that this Calabi-Yau fourfold $ Y_4$ also admits an elliptically fibered K3 fibration. In fact setting the $c_{n,m}$ to constants, i.e.~fixing a point $\underline u_0$, one finds the equation of a K3 surface. The toric base $ B_3$ itself is a $ \mathbb{P}^1 $-fibration with coordinates $ [v:w] $ over $ \mathbb{P}^2 $ with coordinates $ [u_1:u_2:u_3] $.

The Hodge-numbers of $ Y_4$ can be computed by standard techniques to be
\begin{align}
 h^{1,1}( Y_4) = 3, \quad
 h^{2,1}( Y_4) = 1, \quad 
 h^{3,1}( Y_4) = 4358\ . 
\end{align}
Therefore, we find that the smooth hypersurface $ Y_4$ has exactly one $(2,1)$-form.

In our example \eqref{poly_1} the point $\nu^\ast _7$ is the only inner point of a two-dimensional face $ \theta^\ast $ and hence induces the $(2,1)$-form. To see this in more detail, we consider the toric divisor $D_7$ of $\hat \cA_5 $
associated to this inner point. Using the coordinates introduced in \eqref{poly_1} it corresponds to setting $v=0$. Restricted to the hypersurface $ p_\Delta = 0$, i.e.~to $ D^\prime _7 $ and using the scaling relation $\ell_3$ to set $ w = 1$ one thus finds 
\beq \label{ptheta} 
   p_\theta  = y^2 + x^3 + \hat c_1 x y z + \hat c_2 x^2 z^2 + \hat c_3 y z^3 +\hat c_4 x z^4 + \hat c_6 z^6= 0\ ,
\eeq 
where $\hat c_n = a_n(\underline u, v= 0, w=1) = c_{n,0}$ are constant on $ Y_4$, but nevertheless depend on the complex structure moduli. Note that this is simply the equation of a two-torus in Tate form.\footnote{It can be always brought into Weierstrass form $y^2 + x^3+fx + g=0$ as we recall below.} 
This implies that the divisor $D^\prime _7$ is a product of this $R\simeq T^2$ with an $ E = \bbP^2$ 
parameterized by $(u_1,u_2,u_3)$, since $ E$ is fibered over $ R $ and $ R $ is the elliptic fiber fibered over $ E$. The latter exists since the coordinates $\underline{u}$ are unconstrained by \eqref{ptheta} and the $\ell_2$ scaling relation remains a symmetry. It is easy to see from the toric data \eqref{poly_1} that the blow-up by $ \nu^\ast _8 $ separates the cone spanned by $ \nu^\ast _3, \nu^\ast _7 $ and hence resolves the $ \mathbb{C}^4 / \mathbb{Z}_2 $-singularities.
This implies that the divisor $D^\prime _7$ has two non-trivial five-forms build out of the one-forms of the $T^2$ and the volume-form of $ \mathbb{P}^2 $ . In complex coordinates one finds a single $(1,0)$-form on $D^\prime_7$ arising from $ R \simeq T^2 $.

\subsubsection{Picard-Fuchs equations for the three-form periods} \label{subsec:PF_example1}

Let us apply the theory we introduced before in \autoref{subsec:weighted_periods}, to obtain the Picard-Fuchs equations and gain insight in the behavior of the normalized period matrix $ f_{ab} $, that will appear in the effective F-theory action.
We use this section to show how to apply the toric techniques we developed in \autoref{sec:three-forms_hyper} in a simple explicit example.

It is clear from the equation of $ p_\theta $ given in \eqref{ptheta} that the homogeneous coordinate ring and the chiral 
ring of $ \cA_2$, the ambient space of the curve along which we found the $ \mathbb{C}^3/\bbZ_3 $-singularities in the hypersurface $ Y_4 ^{ \rm sing} $, is given by
\begin{equation}
 S_2  = \mathbb{C}[x,y,z] \, , \quad \cR _\theta = S_2  / p _\theta \, ,
\end{equation}
with $ x,y,z $ having the grading $ 2,3,1 $, i.e.~$\cA_2= \mathbb{P}^2(2,3,1)$. Therefore, we find that via the Poincar\'e residue construction
\begin{equation}
 H^{1,0}(R) \simeq \cR_\theta (0) \, , \quad H^{0,1}(R) \simeq \cR_\theta (6)
\end{equation}
are both one-dimensional and generated by
\begin{equation}
 \gamma = \int_\Gamma \frac{1}{p _\theta} \, d\omega_{\cA _2} \,  \in H^{1,0}(R) \, ,
\end{equation}
and its derivative with respect to the one independent complex structure modulus. The holomorphic volume-form of $ \cA_2$ 
is obtained from \eqref{domegaA2_W} to be 
\begin{equation}
 d\omega_{\cA_2} = z dx \wedge dy - x dy \wedge dz + y dx \wedge dz \, .
\end{equation}
For $ p_\theta $ we take the deformation (there are several equivalent choices which differ only in reparametrization) in the 
Weierstrass form \eqref{fg-bdecomp} that allows a comparison to the weak coupling description of the next section
\begin{equation} \label{Fermat_ex1}
 p_\theta = y^2 + x^3 + z^6 + a x z^4
\end{equation}
where $ a = f $ is the only modulus and we take the parameter $ g = 1$. Their derivatives are
\begin{align}
 \partial_a \gamma &= - \int_\Gamma \frac{x z^4}{(p_\theta) ^2} \, d\omega_{\cA_2 } \, \in H^{0,1}(R) \, ,\\
 \partial_a ^2 \gamma &= 2\int_\Gamma \frac{x^2 z^8}{(p_\theta) ^3} \, d\omega_{\cA_2 } \, \in H^1(R,\mathbb{C}) \, ,
\end{align}
and we use the relation
\begin{equation}
 (27 + 4 a^3) x^2 z^8 = 9 z^8 \partial_x p_\theta + (-\frac{3}{2} az^7 + a^2 z^5 x) \partial_z p_\theta \, ,
\end{equation}
to find the Picard-Fuchs equation of $ \gamma $ around the vacuum with $ a = 0 $ to be
\begin{equation} \label{PF_example1}
 (27 + 4 a^3) \gamma^{\prime \prime} + \frac{7}{4}a \gamma + 12 a^2 \gamma^\prime = 0 \, .
\end{equation}
To solve \eqref{PF_example1} we can use the techniques explained in \cite{Lerche:1991wm} combined with the boundary conditions derived in \cite{Greiner:2015mdm}. We know, as for example reviewed in \cite{Denef:2008wq}, that
\begin{equation}
 j(i \hat f(a)) = \frac{4(24a)^3}{\Delta} \, , \quad \Delta = 27 + 4a^3
\end{equation}
and close to the three distinct zeroes $ a_i = 3/4^{1/3}\, \xi^i $ with $ \xi^3 = 1 $ of $ \Delta = 0 $ we find
\begin{equation}
 i\hat f(a) \sim \frac{1}{2\pi i} \text{log}(a - a_i)
\end{equation}
up to $ SL(2,\mathbb{Z}) $-transformations. The boundary conditions derived in \cite{Greiner:2015mdm} are here trivially satisfied, $ \hat f = i \tau $, since the genus of the Riemann surface is one, and hence the coefficient of the linear term is the triple intersection number of the one blow-up divisor in the mirror geometry. Due to the fact that the mirror is also smooth, this number is one. Another way to interpret this result stems from Seiberg-Witten theory, like reviewed in \cite{Lerche:1996xu}. There the exact coupling of an $ SU(2) $ gauge theory was calculated using an elliptic curve  and we find here the same result as a coupling of scalars. The three singularities $ a_i $ can be used as points around which we can expand the period-matrix $ f $ and these three coordinate patches couple the full moduli space of the gauge theory. However, two of these $ a_i $ describe in SW language points of gauge enhancement. In contrast to this, we expand around the large complex structure point of the Calabi-Yau fourfold $ Y_4 $ after transforming to the proper complex structure coordinates $ z^\cK $. In the SW theory this corresponds to the solution at infinity in moduli space, i.e.~deep in the Coulomb branch of the gauge theory.

We have found that $ p_\theta $ is the equation for the elliptic fiber $ R$ over the divisor $ v = 0 $ in the base. This implies in particular, that $ p_\theta $ defines the complex structure $ \tau|_{v=0} $ of the elliptic fiber $ R$ over this divisor. This is defined such that up to 
$ SL(2,\mathbb{Z}) $-transformations we have a holomorphic one-form
\begin{equation} \label{elliptic_gamma}
 \gamma = \hat \alpha + \tau \hat \beta \, \in H^{1,0}(R) \, ,
\end{equation}
for $ \hat \alpha $, $ \hat \beta $ a canonical basis of $ H^1(R ,\mathbb{Z}) $ as introduced in \autoref{Riemann_periods}. This $ \tau $ is the axio-dilaton of Type IIB string theory varying over the base $ B_3$. The important observation here is that 
$ \tau|_{v=0} $ is constant along the divisor $v=0$ in $B_3$, i.e.~does not depend on the base coordinates, but does vary non-trivially 
with the complex structure moduli. To see this, we evaluate 
\begin{equation}
 j(\tau) \big|_{v=0} = \frac{4(24f)^3}{27g^2 + 4f^3} \Big|_{v=0} = C(\hat c_n) \, .
\end{equation}
In order to do that we determine $ f|_{v=0}$, $g|_{v=0} $ using \eqref{fg-bdecomp}, \eqref{bi-def} with the $a_n|_{v=0}$ determined from 
$ p_\theta $ given in \eqref{ptheta}. The result is a non-trivial function of the coefficients $\hat c_n$ of $p_\theta$, these are constants on $Y_4$, but do depend on the complex structure moduli $z^\cK$ of $Y_4$. Note that there are $4358$ such complex structure 
moduli and we will not attempt to find the precise map to the five coefficients $\hat c_n$. Putting everything together, we can thus use $ \tau|_{v=0} $ as normalized period matrix of the curve $ R $ that induces the non-trivial three-forms in the fourfold $ Y_4$. Therefore, we have just shown that
\begin{equation}
 \hat f(z) = i \tau|_{v=0}(\hat c_n) \, ,
\end{equation}
on the full complex structure moduli space of the Calabi-Yau fourfold.

\subsubsection{Weak string coupling limit: a model with two-form moduli}

We next examine the weak string coupling limit of the geometry introduced in subsection \ref{toric_data_example1}. Using Sen's general procedure described in subsection \ref{sen_limit} we add an additional coordinate $ \xi $ to the homogeneous 
coordinate ring of the base $B_3$. The scaling weights of $\xi$ are the degrees of the polynomials associated to the 
anti-canonical bundle $-K_{B_3}$, i.e.~$ \xi $ has the degree of twice the anti-canonical class in the homogeneous 
coordinate ring of $ \hat \cA_4$. Therefore, we find $ Y_3 $ as the Calabi-Yau hypersurface obtained as the blow-up 
of the singular hypersurface $ Y_3 ^{\rm sing} = \mathbb{P}^4[1,1,1,3,6] (12)$. 
Recalling that $ B_3$ is a $ \mathbb{P}^1 $-fibration over $ \mathbb{P}^2 $, the double-cover $ Y_3 $ turns out to be the 
double-cover of $ \mathbb{P}^1 $ fibered over $ \mathbb{P}^2 $. The double-cover of the $\bbP^1$-fiber is a two-torus, or rather an elliptic curve,  $ \mathbb{P}^2[1,1,2](4)$. 

To make this more explicit we again use a toric description. The fan of the ambient space for the three-fold is given by the cones generated by the rays through the points
\begin{equation}
\begin{tabular}[0]{|c  r  r  r  r  r   | c | c | c |}
\hline
  \multicolumn{6}{|c|}{ \text{Example 1:\ \ Toric data of $ \cA_4 $ }}   &\rule[-.2cm]{0cm}{.7cm} coords & $ \ell_1 $ & $ \ell_2 $ \\
  \hline
    $ \nu^\ast_3 = $ & $ ( $ & $ 0$ & $ 0 $ & $ 1 $ &  $ 0 ) $ & $ z_3 = w $   & $ 1 $ & $ 0 $ \\	
    $ \nu^\ast_4 = $ & $ ( $ & $ 0$ & $ 0 $ & $ 0 $ &  $ 1 ) $ & $ z_4 = \xi $ & $ 2 $ & $ 0 $ \\
    $ \nu^\ast_6 = $ & $ ( $ & $ 0$ & $ 0 $ & $-1 $ &  $-2 ) $ & $ z_6 = v $   & $ 1 $ & $-3 $ \\
    \hline
    $ \nu^\ast_1 = $ & $ ( $ & $ 1$ & $ 0 $ & $ 0 $ &  $ 0 ) $ & $ z_1 = u_1 $ & $ 0 $ & $ 1 $ \\
    $ \nu^\ast_2 = $ & $ ( $ & $ 0$ & $ 1 $ & $ 0 $ &  $ 0 ) $ & $ z_2 = u_2 $ & $ 0 $ & $ 1 $ \\
    $ \nu^\ast_5 = $ & $ ( $ & $-1$ & $-1 $ & $-3 $ &  $-6 ) $ & $ z_5 = u_3 $ & $ 0 $ & $ 1 $	\\ 
    \hline
\end{tabular} 
\end{equation}
The hypersurface equation is then denoted by $Q=0$ and from subsection \ref{sen_limit} we can deduce that it has the form
\begin{equation}
 Q= \xi^2 - b_2(\underline u, v,w)
\end{equation}
in the fully blown-up ambient space with
\begin{equation}
 b_2 = a_1 ^2 + 4 a_2 
\end{equation}
specified by the Weierstrass-form of the corresponding fourfold in \eqref{a_example_1}.

One computes the Hodge-numbers to be
\beq
h^{1,1}( Y_3) = 3, \quad h^{2,1}( Y_3)  = 165 \, .
\eeq 
This example was already discussed in the context of mirror symmetry in \cite{Hosono:1993qy}. The resulting threefold is an elliptic fibration over $\bbP^2$ with two sections. It should be stressed that despite the fact that $h^{1,1}(Y_3)=3$ the toric ambient space only admits two non-trivial divisor classes. In fact, we will discuss in the 
following that this can be traced back to the fact that the divisor $v=0$ yields two disjoint $ \mathbb{P}^2 $ when intersected with the hypersurface constraint. These are the two sections, i.e.~two copies of the base. This is also noted in \cite{Gao:2013pra}, 
where a classification of orientifold involutions suitable for Type IIB orientifold compactifications is presented.

To make this more precise, let us analyze the singularities of $ Y_3 ^{ \rm sing} = \mathbb{P}^4[1,1,1,3,6] (12)$ and their resolutions via blow-ups further. 
The ambient space $ \cA_4 = \mathbb{P}^4[1,1,1,3,6] $ has $ \bbC^3 / \mathbb{Z}_3 $-singularities along a curve $ \mathbb{P}^1 $ given by $ [0:0:0:w:\xi] $. The hypersurface intersects this curve in two points, which are identified as double cover of the  
point of the not yet blown up base $ B_3 ^{ \rm sing} = \mathbb{P}^3[1,1,1,3] $, where 
we find $ \mathbb{C}^3/\mathbb{Z}^3 $-singularities. Blowing up this curve of singularities in the ambient space 
by adding $ \nu^\ast _6 $ leads to an exceptional divisor $v=0$, which is a $ \mathbb{P}^2 $ fibration over two points 
of the hypersurface. 
On the hypersurface $ Y_3$ we find that the ambient space divisor $v=0$ splits into two parts
\begin{equation}
 D_6 ^\prime = \{ v=0, \, Q^{(1)} = 0 \} \sim \mathbb{P}^2 \sqcup \mathbb{P}^2
\end{equation}
with coordinates $ [u_1,u_2,u_3,v=0,w, \pm \sqrt{c}w^2] $. Note that $ c $ is a constant, but depends on complex structure moduli. It is given by
\begin{equation} \label{c_def}
 c = b_2 |_{v=0} = c_{1,0}^2 + 4 c_{2,0} \, .
\end{equation}
It obviously measures the separation between the two $ \mathbb{P}^2 $ in which $ D_6 $ splits when intersecting the threefold hypersurface. For $ \hat c_{2,0} = 0 $ we find that $ c $ is a perfect square.

We next investigate the action of the orientifold involution $\sigma: \xi \rightarrow -\xi$. 
From the coordinate description of $ D_6 ^\prime $ we find that the two disjoint $ \mathbb{P}^2 $ are interchanged by the involution $ \sigma $. Therefore, we introduce the two non-toric holomorphic divisors $ D_{6,1} ^\prime $ and $ D_{6,2} ^\prime $ that are the two disjoint $ \mathbb{P}^2 $ such that
$D_6 ^\prime = D_{6,1} ^\prime + D_{6,2} ^\prime$ and $\sigma^\ast (D_{6,1} ^\prime) = D_{6,2} ^\prime$.
It is now straightforward to define an eigenbasis for the involution $ \sigma $ as
\begin{align} \label{Kminusplus}
 K^+ _1 = D_4 ^\prime \, , \quad K^+ _2 = D_6 ^\prime\, , \quad K^- = D_{6,1} ^\prime - D_{6,2} ^\prime \, .
\end{align}
Therefore, we conclude that 
\beq 
h^{1,1} _+ (Y_3 ) = 2 , \qquad h^{1,1} _- (Y_3) = 1\ , 
\eeq
which shows that there is one negative two-from which yields zero-modes for the R-R and NS-NS two-forms
of Type IIB supergravity. Furthermore, we can evaluate the intersection ring to be
\begin{equation}
 I_{Y_3} = 18 (D_6 ^\prime )^3 + 144 (D_4 ^\prime ) ^3 = 18 (D_6 ^\prime) ^3 - 6 D_1 ^\prime (D_6 ^\prime )^2 + 2 (D_1 ^\prime )^2 D_6 ^\prime
\end{equation}
Note that $ D_{6,1} ^\prime \cap D_{6,2} ^\prime = \emptyset $. Due to the symmetry between the components of $ D_6 ^\prime$ and $ D_4 ^\prime$ being exactly the fixed point of this symmetry, we find that the intersections of $ K^- $ appearing linearly vanish. We learn that $ (D_{6,1} ^\prime)^3 = (D_{6,2} ^\prime) ^3 = 9 $, $ (D_{6,1} ^\prime )^2 D_4 ^\prime= (D_{6,1} ^\prime )^2 D_4 ^\prime = 0 $ and $ D_{6,1} ^\prime (D_4 ^\prime)^2 = D_{6,2} ^\prime (D_4 ^\prime)^2 = 0$.\\

From this analysis we see that  all toric divisors are invariant under the involution $ \sigma $. Therefore, we can choose the divisor basis of the base $ B_3 = \hat{\mathbb{P}}^3[1,1,1,3] $ obtained from $ \hat \cA _4 = \hat{\mathbb{P}}^4[1,1,1,3,6] $ by setting $ \xi = 0 $ . This corresponds on the lattice level to projecting to $ \mathbb{Z}^3 $, i.e.~dropping the fourth coordinate of every vertex.

\begin{equation}
\begin{tabular}[0]{|c  r  r  r  r    | c | c | c |}
\hline
  \multicolumn{5}{|c|}{\text{Toric data of $ B_3 $}}  &\rule[-.2cm]{0cm}{.7cm} coords & $ \ell_1 $ & $ \ell_2 $ \\
  \hline
    $ \nu^\ast_3 = $ & $ ( $ & $ 0$ & $ 0 $ & $ 1 ) $ & $ z_3 = w $   & $ 1 $ & $ 0 $ \\	
    $ \nu^\ast_6 = $ & $ ( $ & $ 0$ & $ 0 $ & $-1 ) $ & $ z_6 = v $   & $ 1 $ & $-3 $ \\
    \hline
    $ \nu^\ast_1 = $ & $ ( $ & $ 1$ & $ 0 $ & $ 0 ) $ & $ z_1 = u_1 $ & $ 0 $ & $ 1 $ \\
    $ \nu^\ast_2 = $ & $ ( $ & $ 0$ & $ 1 $ & $ 0 ) $ & $ z_2 = u_2 $ & $ 0 $ & $ 1 $ \\
    $ \nu^\ast_5 = $ & $ ( $ & $-1$ & $-1 $ & $-3 ) $ & $ z_5 = u_3 $ & $ 0 $ & $ 1 $ \\
    \hline
\end{tabular}
\end{equation}
As a consequence, we can use $ D_6 $ and $ D_1 $ as a basis for the divisors on $ B_3 $. For $ Y_3 $ we can choose the corresponding basis via $ D_4 ^\prime = 2 D_6 ^\prime + 6 D_5 ^\prime$ and find
\begin{equation}
 I_{B_3} = 9 D_6 ^3 - 3 D_1 D_6 ^2 + D_1 ^2 D_6 = \frac{1}{2}(18 D_6 ^3 - 6 D_1 D_6 ^2 + 2 D_1 ^2 D_6) \sim \frac{1}{2}I_{Y_3} \, .
\end{equation}
This fits the fact that $ Y_3 $ double-covers $ B_3$ and $ D_{6,1} ^\prime $ and $ D_{6,2} ^\prime $ project down to the same $ \mathbb{P}^2 $ in $ B_3 $.

Let us now discuss what happens to the normalized period matrix $ \hat f = i \tau|_{v=0} $ that we have derived in subsection 
\ref{subsec:PF_example1}, in the weak coupling limit of complex structure space. In this orientifold limit 
the field $\tau_0 = C_0 + i e^{-\phi}$ is actually constant everywhere on $Y_3/\sigma$
and becomes an independent modulus. The identification $\hat f= i \tau_0$ then precisely yields 
the known moduli $N = c- \tau_0 b$ of the orientifold setting, where $c$, $b$ are the zero-modes 
of the R-R and NS-NS two-forms along $K_-$ introduced in \eqref{Kminusplus}.

We close by pointing out that it is important to have $c= \hat c_1 ^2 + 4 \hat c_2 \neq 0$ for this weak coupling analysis to apply. 
Indeed, if we go to the limit $c \rightarrow 0$ we find a spliting of the O7-plane located at $b_2=0$ into $v=0$ and 
$b'_2=0$. Not only would we find intersecting O7-planes, but also the simple identification $\hat f = i\tau_0$ would no longer hold.

\subsection{Example 2: An F-theory model with Wilson line scalars} \label{Wilson_scalars}

In this subsection we construct a second example geometry that we argue to admit Wilson line moduli 
when used as an F-theory background. In this example the three-forms of the Calabi-Yau fourfold stem from a 
genus seven Riemann surface. It turns out that this example features also other interesting properties, such as a 
non-Higgsable gauge group and terminal singularities corresponding to O3-planes.

\subsubsection{Toric data and origin of non-trivial three-forms}

Our starting point is the anti-canonical hypersurface in the weighted projective space $ \cA  _5 = \mathbb{P}^5(1,1,3,3,16,24) $ of degree $ d = 48 $. This space is highly singular, but admits an elliptic fibration necessary to serve as an F-theory background. 
It is easy to see that we have a curve $ R$ along which we find $ \mathbb{C}^3/ \mathbb{Z}_3 $-singularities.
In contrast to the first example this curve $R$ is not the elliptic fiber. It rather arises as a multi-branched cover over 
a $ \mathbb{P}^1 $ of the singular base $ B_3 ^{\rm sing} $.

We can resolve part of the  singularities of the ambient-space $\cA_5$ by moving to a toric space $ \hat \cA_5$ whose fan is obtained by the maximal subdivision of the polyhedron $ \Delta^\ast  $ of $\cA_5$:
\begin{equation} \label{poly_2}
\begin{tabular}[0]{|c  r  r  r  r  r  r  | c | c | c | c | c |}
\hline
 \multicolumn{7}{|c|}{\text{Example 2:\ \ Toric data of $ \hat \cA_5 $}}   & \rule[-.2cm]{0cm}{.7cm} coords & $ F $ & $ \mathbb{P}^2 $ & $ B $ & $ E $\\
  \hline
    $ \nu^\ast_1 = $ & $ ( $ & $ 1$ & $ 0 $ & $ 0 $  & $ 0 $ & $ 0 ) $   & $ z_1 = w   	$ & $ 0 $ & $ 0 $  & $ 1 $ & $ 1 $ \\	
    $ \nu^\ast_2 = $ & $ ( $ & $ 0$ & $ 1 $ & $ 0 $  & $ 0 $ & $ 0 ) $   & $ z_2 = u_1 	$ & $ 0 $  & $ 1 $  & $ 0 $ & $ 0 $ \\
    $ \nu^\ast_3 = $ & $ ( $ & $ 0$ & $ 0 $ & $ 1 $  & $ 0 $ & $ 0 ) $   & $ z_3 = u_2 	$ & $ 0 $  & $ 1 $  & $ 0 $ & $ 0 $ \\
    $ \nu^\ast_4 = $ & $ ( $ & $ 0$ & $ 0 $ & $ 0 $  & $ 1 $ & $ 0 ) $   & $ z_4 = x   	$ & $ 2 $  & $ 0 $  & $ 0 $ & $ 1 $ \\
    $ \nu^\ast_5 = $ & $ ( $ & $ 0$ & $ 0 $ & $ 0 $  & $ 0 $ & $ 1 ) $   & $ z_5 = y   	$ & $ 3 $  & $ 0 $  & $ 0 $ & $ 0 $ \\
    $ \nu^\ast_6 = $ & $ ( $ & $-1$ & $-3 $ & $-3 $  & $-16$ & $-24) $   & $ z_6 = v   	$ & $ 0 $ & $ 0 $  & $ 1 $ & $ 1 $ \\
    $ \nu^\ast_7 = $ & $ ( $ & $ 0$ & $-1 $ & $-1 $ & $ -5 $ & $ -8) $   & $ z_7 = e 	$ & $ 0 $  & $ 0 $  & $ 0 $ & $ -3 $ \\ 
    $ \nu^\ast_8 = $ & $ ( $ & $ 0$ & $-1 $ & $-1 $  & $-6 $ & $ -9) $   & $ z_8 = u_3   $ & $ 0 $  & $ 1 $  & $-3 $ & $ 0 $ \\
    $ \nu^\ast_9 = $ & $ ( $ & $ 0$ & $ 0 $ & $ 0 $  & $-2 $ & $ -3) $   & $ z_9 = z   	$ & $ 1 $  & $-3 $  & $ 1 $ & $ 0 $ \\
    \hline
\end{tabular}
\end{equation}
Note already at this point, that the new ambient space $\hat  \cA_5  $ still contains singularities of the form
\begin{equation}
 \mathbb{C}^4 / \mathbb{Z}_2 \, : \quad (v,w,u_3,y)\quad \rightarrow \quad (-v,-w,-u_3,-y)
\end{equation}
and hence the hypersurface inherits singular points that do not allow for any crepant resolution as pointed out in \cite{Aspinwall:1994ev}. This can be related to the presence of O3-planes. \footnote{ Various aspects of O3-planes have been discussed recently for example in \cite{Collinucci:2010gz,Garcia-Etxebarria:2015wns}}

A number of intriguing features of this model arises due to the geometry of the base $ B_3$. It arises as a non-crepant blow-up of the weighted projective space $ B_3 ^{\rm sing} = \mathbb{P}^3(1,1,3,3) $ with toric data given by
\begin{equation}
\begin{tabular}[0]{|c  r  r  r  r    | c | c | c |}
\hline
 \multicolumn{5}{|c|}{\text{Toric data of $ B_3 $}}    & \rule[-.2cm]{0cm}{.7cm} coords & $ \mathbb{P}^1 $ & $ \mathbb{P}^2 $ \\
  \hline
    $ \nu^\ast_1 = $ & $ ( $ & $ 1$ & $ 0 $ & $ 0 ) $ & $ z_1 = \tilde v $   & $ 1 $ & $ 0 $ \\
    $ \nu^\ast_2 = $ & $ ( $ & $ 0$ & $ 1 $ & $ 0 ) $ & $ z_2 = \tilde u_1 $ & $ 0 $ & $ 1 $ \\
    $ \nu^\ast_3 = $ & $ ( $ & $ 0$ & $ 0 $ & $ 1 ) $ & $ z_3 = \tilde u_2 $ & $ 0 $ & $ 0 $ \\
    $ \nu^\ast_4 = $ & $ ( $ & $-1$ & $-3 $ & $-3 ) $ & $ z_5 = \tilde w $   & $ 1 $ & $ 1 $ \\
    $ \nu^\ast_5 = $ & $ ( $ & $ 0$ & $-1 $ & $-1 ) $ & $ z_6 = \tilde u_3 $ & $-3 $ & $ 1 $ \\
    \hline
\end{tabular} \, .
\end{equation}
It can be interpreted as a generalization of a Hirzebruch surface, i.e.~a $ \mathbb{P}^2 $-fibration over $ \mathbb{P}^1 $. We note in particular, that the point $ \nu^\ast _5 $ does lie in the interior of the convex hull of the remaining points and correspondingly the new polyhedron is no longer convex. The consequence is that the anti-canonical bundle $ -K_{B_3 } $ of the base has only global sections that vanish over the locus $ \{ \tilde u_3 = 0 \} \simeq \mathbb{P}^1 \times \mathbb{P}^1 $, i.e.~$ -K_{B_3 } $ is not ample. In the F-theory picture this will lead to a non-Higgsable cluster as described in \cite{Grassi:2014zxa,Morrison:2014lca}, i.e.~to the generic existence of a non-Abelian gauge group in this setting. The base $ B_3 $ has been analyzed recently in detail in \cite{Berglund:2016nvh}.

The ambient space $\hat \cA _5 $ has the fibration structure given by the projection $ \pi: \, \hat \cA_5 \, \rightarrow \, B_3 $, which reads in homogeneous coordinates
\begin{equation}
 \pi: \, [v:w:u_1:u_2:u_3:x:y:z:e] \, \mapsto \, [\tilde v = v: \tilde w = w: \tilde u_1 = u_1: \tilde u_2 = u_2: \tilde u_3 = e u_3] \, .
\end{equation}

Due to the non-Higgsable gauge group, $ Y_4 $ can only be written in Tate form after blowing down the exceptional divisor $e=0$, i.e.~setting $ e=1 $:
\beq \label{p2}
 p_{\Delta} = y^2 + e x^3 + \hat a_1\, x y + \hat a_2 \, x^2 +\hat a_3 \, y + \hat a_4 \, x  + \hat a_6 = 0\ ,
\eeq
with $ \hat a_i $ global sections of $ K_{B_3} ^{-i} $. 
Due to the properties of $ K_{B_3} ^{-1} $ these $ \hat a_n $ have common factors of $ u_3 e = \tilde u_3 $ 
independently of the point in complex structure space. This shows that the non-Higgsable cluster with the 
enhanced gauge group is located on the divisor $ \tilde u_3 = 0 $ in the base. The 
singularity type can be easily read of by translating \eqref{p2} into Weierstrass form using \eqref{fg-bdecomp}, \eqref{bi-def}.
We then obtain a singularity of orders $ (2,2,4) = (f,g,\Delta) $, where $ \Delta $ is the discriminant as above. 
This leads to a type $ IV $ singularity and the exact gauge group, which is either $Sp(1)$ or $SU(3)$, 
can be derived from monodromy considerations as we recall below.
The generic anti-canonical hypersurface $ Y_4$ of the ambient space $ \hat \cA_5$ has Hodge numbers
\beq
   h^{1,1}( Y_4) = 4, \quad h^{2,1}( Y_4) = 7, \quad h^{3,1}( Y_4) = 3443 , \quad h^{2,2}( Y_4) = 13818 \, .
\eeq
This implies that $Y_4$ indeed has seven $(2,1)$-forms and we claim that these arise from a single Riemann surface of 
genus $ g = 7 $.

There is only one two-dimensional face $ \theta^\ast $ of the polyhedron spanned by $ \nu^\ast _1, \nu^\ast _4, \nu^\ast _6 $ that contains an interior integral point. This interior point is $ \nu^\ast _7 $ and we add this point to resolve the $ \mathbb{C}^3 /\mathbb{Z}_3 $-singularity along the surface $ \cA_2 = \mathbb{P}^2(1,1,8) $ given as the subspace of $ \cA_5 $ with $ w=v=x=0 $. The anti-canonical hypersurface $ Y_4$ intersects $ \cA_2$ in a Riemann surface $ R $ given by
\begin{equation}
 R = \mathbb{P}^2(1,1,8)[16],  \quad g = 7 \, .
\end{equation}
This can also be seen from the dual face $ \theta $ whose inner points correspond to the monomials
\begin{equation}
 p^\prime _a = u_1 ^a u_2 ^{6-a} \, \in \cR_\theta (6) , \quad a = 0, \ldots, 6
\end{equation}
where we already divided out the common factor $ u_1 u_2 y $ as described in \autoref{Riemann_periods}. The exceptional divisor resolving this singularity is a fibration over $ R$ with fiber $ E = \mathbb{P}^2(1,1,16) $.

Expanding the Weiserstrass form of $Y_4$ around the singular divisor $ D_e = \{ e = 0 \} $, we find
\begin{equation}
 g = g_2 e ^2 + \cO(e ^3) \, , \quad g_2 = g_2(u_1,u_2)
\end{equation}
and this $ g_2 $ is precisely the degree $ 16 $ polynomial in $ u_1, u_2 $ defining the Riemann surface $ R  $ by
\begin{equation} \label{p_theta_ex2}
 R : \quad p _\theta = y^2 - g_2  = 0\, .
\end{equation}
The resulting gauge group over $ D_3 = \{ \tilde u_3 = 0 \} $ in $ B_3$ is $ Sp(1) $ for general $ g_2 $ and if $ g_2 = \gamma^2 $, i.e.~for $ g_2 $ a perfect square, we have an enhancement to $ SU(3) $.

\subsubsection{Comments on the weak string coupling limit}

So what happens to this curve in the weak coupling limit? 
For a $ IV $ singularity, there should be no straightforward perturbative limit in which 
$ \tau $ can be made constant and $\text{Im}\,\tau$ can be made very large over the base.
The general hypersurface equation derived from the naive Sen limit is
\begin{equation}
 Q=\xi^2 - b_2 = \xi^2 - \tilde u_3 \cdot b^\prime _2 = 0\, ,
\end{equation}
implying that the O7-plane splits in two intersecting branches, $ \tilde u_3 = 0 $ and $ b^\prime _2 = 0 $. At the intersection of the 
O7-planes perturbative string theory breaks down and hence there is no weak coupling description. However, we can still try to learn 
some of the aspects of the D7-branes in this setting. 

In fact, in the following we want to connect the curve \eqref{p_theta_ex2}
and Wilson line moduli located on D7-branes. As explained in \cite{Jockers:2004yj} 
the number of Wilson line moduli arising from a D7-brane image-D7-brane on a divisor $S \cup \sigma(S)$ 
of the threefold $ Y_3 $ is given by
\begin{equation}
 \text{Number of Wilson line moduli on } S \, : \quad h^{1,0} _- (S\cup \sigma(S)) \, .
\end{equation}
These are the $(1,0)$-forms on the union of $ S $ and its image that get 
projected out when considering the orientifold quotient. Therefore, we suggest that the Wilson lines 
arise in $ S \cup \sigma(S) $ as arcs in $ S $ that connect two components of $ S \cap \sigma(S)$. These arcs close to one-cycles in $ S \cup \sigma(S) $, but get projected out when we take the quotient $ Y_3/ \sigma = B_3$. Note here that $ S \cap \sigma(S) $ is equal to $ \text{O7} \cap S $. In our situation $Y_3$ is still a fibration over $ \mathbb{P}^1 $ with coordinates $ [v:w] $ and hence this will also hold for $ S \cap \sigma(S) $, 
i.e.~we suggest that $ S \cap \sigma(S) $ is a covering space of the base $ \mathbb{P}^1 $ given by 
\begin{equation}
 S \cap \sigma(S) = \{ \xi=0, \tilde u_3=0, g_2 = 0 \} \subset Y_3 \, ,
\end{equation}
where $ \xi = \tilde u_3 = 0 $ is the location of one branch of the O7-plane in $ Y_3 $. We also note that the divisor inducing the three-forms in the fourfold projects down to the $ \tilde u_3 =0 $ divisor of $B_3$. Recall that the locations of the seven-branes in a 
general F-theory model are given by the zeroes of the discriminant $ \Delta $. We can expand $\Delta$ around $ \tilde u_3 = 0 $ to
\begin{equation}
 \Delta \approx b_2 ^2 (b_2b_6 - b_4 ^2) = \tilde u_3 ^5 (b^\prime _2 )^3 g_2 + \cO(\tilde u_3 ^6) \, .
\end{equation}
This implies that in the weak coupling limit $ g_2 $ describes the intersection of the D7-brane  
in the form of a Whitney-Umbrella explained in \cite{Collinucci:2008pf} with the O7-branch given by $ \tilde u_3 = 0 $. 
For our considerations, it is just important that a D7-brane is path connected, 
but the shape away from the O7-plane is irrelevant for our analysis of Wilson lines. 
Therefore, we find that
\begin{equation}
 S \cap \sigma(S) = \bigcup_{i=1} ^{16} \big( \{p_i\} \times \mathbb{P}^1 \big) \, , \quad g_2 (p_i) = 0 \, .
\end{equation}

The points $ p_i $ can be interpreted as branching loci of the auxiliary hyperelliptic curve which is given by \eqref{p_theta_ex2}. Hence we find
\begin{equation}
 h^{1,0} _- (S\cup \sigma(S)) = 7 \, .
\end{equation}
Choosing a normalized basis $ \hat \alpha_a, \hat \beta^a $ for the cocycles arising from this procedure we can give a basis for $ H^{1,0} _- (S \cup \sigma(S)) $ as
\begin{equation}
 \gamma_a = \hat \alpha_a + i \hat f_{ab} \hat \beta^b \, \in H^{1,0} _- (S \cup \sigma(S)) \, ,
\end{equation}
with $ \hat f_{ab} $ the normalized period matrix of the curve $ R $ discussed in \autoref{Riemann_periods}.
The coupling of the corresponding fields, the Wilson moduli $ N_\cA = N_a $, is given by the the normalized period matrix $ f_{\cA \cB} = \hat f_{ab} $ of $ R $.

Let us close by making one final observation for this example geometry. We can also resolve the $ \mathbb{Z}_2 $-singular points of the fourfold by blowing-up the ambient space $ \cA _5 $. This requires adding the exterior point
\begin{equation}
 \nu^\ast _{10} = (0,-2,-2,-10,-15) \, .
\end{equation}
This has, however, drastic consequences. As already mentioned before, there is no way to resolve the $ \mathbb{Z}_2 $-singular 
points in a crepant way, i.e.~preserving the anti-canonical bundle of the ambient-space. 
Closer inspection of the blow-up tells us that this blow-up is not crepant, but leads to a Calabi-Yau hypersurface in a new ambient-space 
that has a different triangulation not compatible with the old triangulation structure. 
This leads to a change in topology, which can be seen from the Hodge-numbers
\begin{equation}
 h^{1,1} _{new} = 5, \quad h^{2,1}_{new} = 0 , \quad h^{3,1}_{new} = 3435, \quad \chi = \chi_{old} = 20688\, ,
\end{equation}
with the Euler number $ \chi $ being preserved. This extremal transition between the two fourfolds follows a similar pattern 
as the conifold transition along curves described in \cite{Intriligator:2012ue}. The relations to the non-trivial three-form cohomology 
can also be made precise: the blow-up obstructs precisely the complex structure deformations described by $ g_2 $ setting it to zero and hence also obstructing the three-form cohomology. This obstruction leads to a further gauge-enhancement to $ G_2 $ along $ D_3 $ and also the weak coupling limit is no longer singular, i.e.~the O7-plane does no longer branch.


\section{Concluding remarks and outlook} \label{sec:outlook}

In this work we introduced a framework to explicitly derive the moduli dependence of non-trivial three-forms on 
Calabi-Yau fourfolds. Our focus was on geometries realized as hypersurfaces in toric ambient spaces for which we 
argued that properties of the three-form cohomology are essentially inherited from one-forms on
embedded Riemann surfaces supplemented by topological information about the 
corresponding resolution divisors. We also described concrete example geometries that highlight simple physical applications 
of these concepts. In the following we would like to point out several directions for future research. 

A first interesting direction is to further extend and interpret the calculations outlined in \autoref{sec:three-forms_hyper} in the context of mirror symmetry for Calabi-Yau fourfolds \cite{Greene:1993vm,Mayr:1996sh,Alim:2012gq}. In particular, it would be desirable to derive a general expression 
for the Picard-Fuchs equations for three-form periods in terms of the toric data of the ambient space in analogy to the discussion 
of \cite{Hosono:1993qy}. Furthermore, one striking observation to exploit mirror symmetry can be made by recalling 
the construction of the period matrix of the intermediate Jacobian. We note that mirror symmetry exchanges the two-dimensional faces 
$\theta_\alpha$ with their duals $\theta^\ast _\alpha$ and hence maps the one-forms on the Riemann surface $R_\alpha$ 
to the resolution divisors $D'_{l^\alpha}$. Indeed the number of $(1,0)$-forms, given by $\ell^\prime (\theta_\alpha)$ in \eqref{h21-ells},  
and the number of resolution divisors, given by $\ell'(\theta^\ast _\alpha)$ in \eqref{h21-ells}, are exchanged. 
This implies that the relevant intersection data for the $D'_{l^\alpha}$ must be captured by the period matrix of three-forms on the mirror 
geometry, at least at certain points in complex structure moduli space. Indeed, this behavior was already found around the large volume and large complex structure point in \cite{Greiner:2015mdm}. This observation is further supported by 
basic facts from Landau-Ginzburg orbifolds \cite{Intriligator:1990ua,Vafa:1989xc,Lerche:1989uy}, since in these 
constructions both the intersection data and periods are determined by the structure of the chiral rings of the fourfold and its mirror. 
One can thus conjecture that the complex structure dependent three-form periods calculate on the mirror geometry the K\"ahler moduli dependent quantum corrections to the intersection numbers between integral three-forms and two-forms.
It is then evident to suggest that these K\"ahler moduli corrections already cover world-sheet instanton corrections to the three-form couplings, when using the Calabi-Yau fourfold as a string theory background. It would be very interesting to access these corrections directly on 
the K\"ahler moduli side and establish their physical interpretation.  

A second promising direction for future research is to apply our results in the duality between F-theory and the heterotic string theories. 
The relevance of three-forms in this duality was already pointed out, for example, in \cite{Friedman:1997yq,Curio:1998bva,Diaconescu:1999it}. Indeed, in heterotic compactifications on elliptically fibered Calabi-Yau threefolds with stable vector bundles, the moduli space of certain vector bundle moduli also admits the structure of a Jacobian variety. By duality this Jacobian turns out to be isomorphic to the intermediate Jacobian of the corresponding $ K3 $-fibered Calabi-Yau fourfold. The described powerful techniques 
available for analyzing the three-form periods on fourfolds might help to shed new light on the derivations required in the dual heterotic setting. 
Our first example describes a simple case of such an F-theory compactification with non-trivial intermediate Jacobian for which the comparison to its heterotic dual geometry can be performed explicitly. It is an interesting task to analyze several such dual settings in detail. 

The possibility of a direct calculation of the three-form metric also has immediate applications in string phenomenology. The scalars arising 
from the three-form modes can correspond to scalar fields in an F-theory compactification to four space-time dimensions.  These scalars are naturally axions, since the shift-symmetry is inherited from the forms of the higher-dimensional theory. The axion decay constants are thus given by 
the three-form metric and determines the coupling to the K\"ahler and complex structure moduli and thus can be 
derived explicitly for a given fourfold geometry. Since these geometries might not be at the weak string coupling limit of F-theory, one 
might be lead to uncovered new possibilities for F-theory model building. 
For example, our second example is admitting, if at all, a very complicated weak string coupling limit, but can be 
analyzed nevertheless using the presented geometric techniques. In this example also 
non-Higgsable clusters and O3-planes are present and it is interesting to investigate the 
physics of these objects in the presents of a non-trivial three-form cohomology. 
It is important to stress that consistency of Calabi-Yau fourfold compactifications generically require the inclusion 
of background fluxes \cite{Sethi:1996es}. It is well-known that these are also relevant in 
most phenomenological applications. Therefore, it is of immediate interest to generalize 
our discussion to include background fluxes. 
This will be particularly interesting in singular limits of the geometry, which are relevant in the 
construction of F-theory vacua. In particular, the intermediate 
Jacobian plays an important role in the computation of the spectrum of the effective theory as, 
for example, suggested by the constructions of \cite{Bies:2014sra}.
The generalization 
to include fluxes will also be relevant in discussing extremal transitions in Calabi-Yau 
fourfolds that change the number of three-forms.

To conclude this list of potential future directions, let us also mention the probably most obvious generalization 
of the discussions presented in this work and its immediate relevance for F-theory compactification. 
In fact, in this paper we have only considered hypersurfaces in 
toric ambient spaces. A generalization to complete intersections, i.e.~Calabi-Yau manifolds described by 
more then one equation, would be desirable. This is particularly evident when recalling that in F-theory compactifications 
on elliptically fibered fourfolds, the non-trivial three-form cohomology of the base yields U(1)-gauge fields in 
the four-dimensional effective theory \cite{Grimm:2010ks}. The function $f_{\cA \cB}$ then corresponds to the 
gauge coupling function and it is an interesting task to use geometric techniques for Calabi-Yau fourfolds to study 
setups away from weak coupling.

\subsubsection*{Acknowledgments}
We would like to thank Albrecht Klemm for initial collaboration. We are also grateful to 
Ralph Blumenhagen, Ilka Brunner, Michael Fuchs, I\~naki Garc\'ia-Etxebarria, 
 Albrecht Klemm, Diego Regalado, and Irene Valenzuela for illuminating discussions.
This work was supported in part by a grant of the Max Planck Society.

 \appendix
 
\section{Calabi-Yau fourfold hypersurfaces} \label{hypersurface_appendix}

In this subsection we discuss the explicit construction of Calabi-Yau hypersurfaces with non-trivial 
three-form cohomology by using toric techniques. The key feature here is to generalize the usual discussion of 
Fano toric varieties as ambient spaces to the non-Fano case. In other words, we will considers toric varieties for which the 
anti-canonical bundle is not ample. This requirement is based on the Lefschetz-hyperplane 
theorem that forbids the existence of non-trivial three-forms on a toric Calabi-Yau fourfold hypersurface if the anti-canonical 
bundle of the ambient space is ample. The generalization that we will consider are toric ambient spaces with semiample 
anti-canonical bundle, which, as we will recall, can admit a non-trivial three-form cohomology.

The starting point for the construction of the toric ambient space is a polyhedron $ \Delta^\ast \subset N_\mathbb{Q} $ in the rational extension of the lattice $ N \simeq \mathbb{Z}^5 $.
The polyhedron $ \Delta^\ast $ describes the ambient toric variety $ \cA_5 $, as explained, for example, in \cite{Fulton:1993}. Integral points of the polyhedron $ \Delta^\ast $ will be denoted by $ \nu^\ast _i $ and these define the rays $ \tau_i $ whose span will form cones for the fan $ \Sigma(\Delta^\ast) $ describing the structure of the toric ambient space $ \cA_5 $. In the following we will always assume that all cones are simplicial, i.e.~every cone is a cone over a simplex of the polyhedron. This is not a trivial assumption in higher dimensions, but is for example satisfied by fans for weighted projective spaces. We assume that this can be achieved by a maximal star-subdivison of $ \Delta^\ast $ such that all rays through $ N \cap \Delta^\ast $ are part of the fan $ \Sigma(\Delta^\ast) $. As a result the space $ \cA_5 $ will only have $ \mathbb{Z}_n $-orbifold singularities along subspaces of codimension greater than one.

The hypersurface $ Y_4^{\rm sing}$ that describes the Calabi-Yau fourfold is given by the convex Newton-Polyhedron $ \Delta \subset M_\mathbb{Q} = (N^\ast)_\mathbb{Q} $ whose integral points $ \nu_i $ correspond to the monomials of the polynomial whose zero set is $ Y_4^{\rm sing}$. In a more mathematical language, the convex polyhedron $ \Delta $ describes a class of toric Weil-divisors $ D_\Delta $ that are zero-sections of line-bundles $ L_\Delta $. Varying the coefficients $ a_i $ of the monomials corresponds to a varation of the hypersurface in its divisor class $ D_\Delta $. The details of this construction were nicely described in \cite{Batyrev:1994hm} and there also the singularity structure of resulting algebraic varieties is discussed in detail. To obtain a Calabi-Yau variety from this construction, we need to choose a special divisor of the ambient space $ \cA_5 $, its anti-canonical divisor
\begin{equation}
 D_\Delta = -K_{\cA_5} = \sum_{\nu^\ast _i \in \Delta^\ast} D_i \, ,
\end{equation}
where $ D_i $ is the divisor associated to the ray through $ \nu^\ast _i $. Here the homogeneous coordinate $ X_i $ associated to the ray through $\nu^\ast _i $ vanishes, i.e.~$ D_i = \{ X_i = 0 \} $. The corresponding ring of homogeneous coordinates $ X_i $ as defined in \cite{1992alg.geom.10008C} is given by
\begin{equation}
 S_5 = \mathbb{C}[X_i \, , \, \nu^\ast _i \in \Delta^\ast ] \, .
\end{equation}
This ring has a natural grading by divisor classes $ \alpha \in A_{4}(\cA_5) $, where $ A_4(\cA_5) $ is the set of Weil divisors of $ \cA_5 $ modulo rational equivalence, called Chow group of $ \cA_5 $. A monomial $ f = \prod_i X_i ^{b_i} $ has degree $ \text{deg}(f)=\alpha $, if $ \alpha = [\sum_i b_i D_i] $.

A further necessary condition equivalent to the associated anti-canonical line bundle $ L_\Delta $ being trivial is to demand reflexivity of $ \Delta $, i.e.~$ \Delta $ should have exactly one interior point, that we can always shift to the origin of $ M $. This is equivalent to $ \Delta^\ast $ being reflexive, if both polytopes are convex, and we can also describe $ \Delta $ via
\begin{equation}
 \Delta = (\Delta ^\ast)^\ast = \{ u \in M_{\mathbb{Q}} \, | \, \langle u,v \rangle \geq -1 \, , \forall v \in \Delta^\ast \} \, .
\end{equation}
The corresponding a priori singular hypersurface $ Y_4 ^{\rm sing}$ or rather the global section of $ -K_{\cA_5} $ whose zero locus is $ Y_4 ^{\rm sing} $ is given by 
\begin{equation} \label{p_Delta}
 p_\Delta = \sum_{\nu_j \in \Delta \cap M} a_j \prod_{\nu^\ast _i \in \Delta^\ast \cap N} X_i ^{\langle \nu_j , \nu^\ast _i \rangle + 1} \quad \in S(- K_{\cA_5})
\end{equation}
where we associated to every ray of the triangulation of the polyhedron $ \Delta^\ast $ a homogeneous coordinate $ X_i $. Toric blow-ups of the ambient space $ \cA_5 $ can be performed by adding a homogeneous coordinate for every ray through an integral point of $ N $ with the corresponding change of the triangulation of $ \Delta^\ast $ and therefore also changing the fan of $ \cA_5 $.
If such an integral point is not contained in the boundary of the reflexive $ \Delta^\ast $, we will also change $ \Delta $ by the blow-up and generically change the number of possible deformations corresponding to integral points of $ \Delta $. This is called a non-crepant resolution. We will assume that we can resolve singularities by crepant resolutions, i.e.~preserving the anti-canonical divisor class and hence $ \Delta $. We will also assume that there is a transverse and quasi-smooth hypersurface in the anti-canonical divisor class.
We denote the resolved smooth Calabi-Yau hypersurface by $Y_4$.

Applying to this setting the Lefschetz hyperplane theorem as reviewed in \autoref{sec:three-forms_hyper} we find that $ \Delta $ can not define an ample anti-canonical divisor $ -K_{\cA_5} $ and hence $ \cA_5 $ can not be Fano.
In particular, we find that $ \cK_{\cA_5} $ is not ample, if $ \cA_5 $ supports non-trivial three-forms. The reason for this is, that in toric geometry for an ample Cartier toric divisor over a complete toric variety we have a one-to-one correspondence between vertices of $ \Delta $ and maximal-dimensional cones in $ \Delta^\ast $, see \cite{Fulton:1993} section 3.4 . This is obviously not true for a crepant resolution, i.e.~a resolution obtained from adding a ray through a point in the interior of a face of $ \Delta^\ast $ to the fan. In contrast to the standard works for Calabi-Yau hypersurfaces in Fano toric varieties, e.g. in \cite{Batyrev:1994hm} for threefolds and in \cite{Braun:2014xka} for the sextic fourfold, where the anti-canonical divisor is ample, we have to deal with the case where the anti-canonical divisor is only semiample and hence compatible with the resolution of singularities. This was done in the work of Mavlyutov, for example in \cite{Mavlyutov:2000hw,Mavlyutov:2003fa,2006math.....10422M}. Here the author generalizes the toric formalism to include divisors of the hypersurface that carry themselves non-trivial cohomology and induce additional non-trivial cocycles of the full hypersurface. These divisors corresponds to the exceptional divisors of the blow-ups described above \footnote{It can also be shown by methods derived in \cite{Batyrev:1994pg}, that three-form cohomology on a generic anti-canonical hypersurface in a toric variety always arises from one-forms of toric divisors. For these toric divisors to have non-trivial one-forms it is necessary that the corresponding Newton-polyhedron of its hypersurface equation is two-dimensional, and hence the anti-canonical bundle is not ample on these divisors and hence not ample on the whole ambient space. }.

Let us consider in more detail the resolution of $ Y^{\rm sing}_4 $ to the smooth hypersurface $ Y_4 $ in the resolved 
ambient space $ \hat \cA_5 $. This makes $  Y_4 $ a regular semiample hypersurface in the 
complete simplicial toric variety $ \hat \cA_5 $. 
We denote the toric divisors in $\hat \cA_5$ by $D_l$ and their restriction to $Y_4$ by $D_l'$, i.e.~
\beq
   D_l = \{ \, X_l = 0, \quad \nu_l^* \in \Delta^* \, \}\ , \qquad D_l' = D_l \cap Y_4\ .
\eeq
The inclusion will be denote by $\iota_l : D^\prime _l \hookrightarrow  Y_4 $.
To find the origin of the three-form cohomology classes in $  Y_4 $ we use the exact sequence in equation (7) 
of \cite{1998math.....12163M} which leads to the isomorphism
\begin{equation} \label{one_to_three-form_exact}
 0 \longrightarrow \bigoplus_{\nu^\ast _l} H^1( D^\prime _l, \mathbb{C}) \xrightarrow{ \, \oplus \iota_{l \ast} \, } H^3( Y_4, \mathbb{C}) \longrightarrow 0 \, ,
\end{equation}
where the morphism is the direct sum of Gysin morphisms $ \iota_{l \ast} $ of the inclusions $ \iota_l$. This map is defined using the Hodge star (see e.g.~\cite{Voisin03hodgetheory}). For the geometries under consideration we can translate 
\eqref{one_to_three-form_exact} to 
\begin{equation} \label{five_to_five_form}
 0 \longrightarrow  H^5( Y_4, \mathbb{C}) \xrightarrow{ \, \oplus \iota_l ^\ast \, } \bigoplus_{\nu^\ast _l} H^5( D^\prime _l, \mathbb{C}) \longrightarrow 0 \, ,
\end{equation}
where now the isomorphism is given by the sum of $ \iota_l ^\ast $ restricting a five-form on $  Y_4 $ to the various divisors $ D^\prime _l $. This is in particular compatible with the Hodge-structure on $ Y_4 $. Therefore, we see that all five-forms arise from five-forms of divisors $ D^\prime _l $ on $ Y_4 $ induced by the toric divisors $ D_l $ of the resolved ambient space $ \hat \cA_5 $. Due to Hodge duality both approaches are, however, equivalent, i.e.~if we can find toric divisors with five-forms these divisors will also carry the dual one-forms and vice versa.

This poses the problem to find all divisors among the $\{ D^\prime _l, \, \nu_l^* \in \Delta^*\} $ of $  Y_4 $ that support non-trivial one-forms which we discuss in the next section.

\section{Toric divisors of $ Y_4 $ with non-trivial one-form cohomology} \label{Divisor_appendix}

The divisors of a simplicial toric variety $ \hat \cA_5 $ correspond to the rays through integral points $ \nu^\ast $ in the boundary of the polyhedron $ \Delta^\ast $ and can be classified by the codimension $ \text{codim}(\theta^\ast) $ of the face $ \theta^\ast \subset \Delta^\ast $ such that $ \nu^\ast \in \text{int}(\theta^\ast) \cap N $, as was done in \cite{Klemm:1996ts}. Here we want to be a bit more explicit and focus especially on the origin of the non-trivial five-forms of $  Y_4 $ and hence also the non-trivial three-forms on $  Y_4 $ by the Hodge star isomorphism.

To understand the geometric structure of the divisors $ D_l ^\prime = D_l \cap  Y_4 $, which are again semiample hypersurfaces in the toric variety $ D_l $, we will first review the construction of the $n$-dimensional toric subvarieties $ \cA_n $ of $ \cA_5 $. The subvariety $ \cA_n $ corresponding to an $(4-n)$-dimensional face $ \theta^\ast $ in $ \Delta^\ast \subset N_\mathbb{Q} $ is constructed as follows. The face $ \theta^\ast $ defines an $ (5-n) $-dimensional cone $ \sigma $ in $ N_\mathbb{Q} $ and the new lattices $ M_n, N_n $ are defined as
\begin{align} \label{An_lattices}
 N_n &= N(\sigma) = N/N_\sigma \, , \quad N_\sigma = N \cap \mathbb{Q} \cdot \sigma \subset N \\
 M_n &= M(\sigma) = M \cap \sigma^\perp \, , \nn 
\end{align}
which are both $ n $-dimensional lattices. The fan for $ \cA_n $ is given by the set $ Star(\sigma) $, containing all cones over faces of $ \Delta^\ast $ that share faces with $ \theta^\ast $, projected to $ N(\sigma) $. These faces form again a star subdivison of a polytope $ \Delta^\ast _n $ in $ N(\sigma) $ and $ \sigma $ gets projected to the origin of $ N(\sigma) $.

Correspondingly, the homogeneous coordinate ring for $ \cA_{n,\theta^\ast} $, which we call $ S_{n,\theta^\ast} $, is given by
\begin{equation} \label{S_ndef}
 S_{n,\theta^\ast} = \mathbb{C}[X_i , \, \nu^\ast_i \in \Delta^\ast _n] \subset \mathbb{C}[X_i , \, \nu^\ast_i \in \Delta^\ast] /\langle X_i, \nu^\ast_i \in \theta^\ast \rangle = S_5 /\langle X_i, \nu^\ast_i \in \theta^\ast \rangle \, .
\end{equation}
These rings are generated by the monomials $ \prod_i X_i ^{b_i} $ that are graded by the class $ [\sum_i b_i D_i ] \, \in \, A_{n-1}(\cA_{n,\theta^\ast}) $. There is only an inclusive relation, since there are homogeneous coordinates generating $ S_5 $ corresponding to divisors that do not intersect $ \cA_{n,\theta^\ast} $. These homogeneous coordinates can be set to one for our considerations.

By construction, our polynomial $ p_\Delta $ is in $ S_5(-K_{\cA_5}) $, i.e.~it is in the class of the anti-canonical divisor of $ \cA_5 $. This implies that the restriction to $ \cA_{n,\theta^\ast} $ acts as
\begin{equation} 
 S_5(-K_{\cA_5}) \quad \rightarrow \quad S_{n,\theta^\ast}(-K_{\cA_5} \big|_{\cA_{n,\theta^\ast}}) \quad \Rightarrow \quad p_\Delta \mapsto p_\theta \, ,
\end{equation}
i.e.~we set all homogeneous coordinates $ X_i $ corresponding to $ \nu^\ast _i \in \theta^\ast $ to zero and all homogeneous coordinates of divisors not intersecting $ \cA_{n,\theta^\ast} $ to one.
The monomials of $ p_\theta $, i.e.~the global sections of $ H^0(\cA_5, \cK_{\cA_5}) $ surviving the projection to $ \cA_{n,\theta^\ast} $, correspond to the monomials in the face $ \theta $ dual to $ \theta^\ast $
\begin{equation}
 \theta = \{ v \in \Delta \, | \, \langle v, w \rangle = -1 \, , \, \forall w \in \theta^\ast \} \, .
\end{equation}
This in particular implies that, following \cite{Mavlyutov:2000hw}, the divisors $ D^\prime _l $ are so called 
$\text{dim}(\theta)$-semiample hypersurfaces of the toric varieties $ D_l $. From this it can be deduced that
\begin{equation}
 H^{k,0}(D^\prime _l,\mathbb{C}) = 0 \quad \text{for} \quad 0 < k < \text{dim}(\theta) -1 \, .
\end{equation}
Therefore, we can only have non-trivial three-forms that arise from $ (4-n) = 2 $-semiample divisors and hence from a pair of two-dimensional faces $ (\theta^*, \theta) $. 

From here on we will consider dual pairs of faces that are two-semiample, i.e. $ n = 2 $ in \eqref{An_lattices} and \eqref{S_ndef}.
We denote the relevant faces by 
\beq
  (\theta^*_\alpha,\theta_\alpha)\ , \quad \text{dim}(\theta^\ast _\alpha) = 2\ ,\qquad \alpha = 1,\ldots, n_2\ ,
\eeq
where $n_2$ denotes the number of two-dimensional faces in $\Delta^*$. These faces exist due to the 
blow-up procedure as described above. Thus, we can associate divisors $D _{l_\alpha} $ to 
each pair $(\theta^*_\alpha,\theta_\alpha)$, i.e.~
\beq \label{introduceDl}
   D_{l_\alpha }\ :\quad  \nu^\ast _{l_\alpha} \in \text{int}(\theta^\ast_\alpha) \cap N \ ,\qquad l_{\alpha} = 1,\ldots, \ell'(\theta_\alpha^*)\ ,
\eeq
where $\ell'(\theta_\alpha^*)$ counts the number of divisors satisfying this condition for the face $\theta^*_\alpha$.
 The divisor $ D^\prime _{l_\alpha} = D_{l_\alpha } \cap Y_4 $ can also be written as $ D^\prime _{l_\alpha} = V(\tau_{l_\alpha}) $, where $ \tau_{l_\alpha} $ is the ray through $ \nu^\ast _{l_\alpha} $, and admits a fibration structure. If $ \nu^\ast _{l_\alpha} $ is contained in the interior of an two-dimensional face $ \theta^\ast _\alpha $ and hence $ \tau_{l_\alpha} $ is contained in the interior of the three-dimensional cone $ \sigma_\alpha $ we find that the polyhedron $ \Delta^\ast _4 $ for $ V(\tau_{l_\alpha}) $ is given by the projection of $ \Delta^\ast $ to $ N(\tau_{l_\alpha}) $ and hence has the image of $ \theta^\ast _\alpha $ as subpolyhedron. Correspondingly we find the fibration-structure for $ D_{l_\alpha} $
\begin{equation} \label{Elalph_fibrD}
 \begin{tikzpicture}
  \matrix (m) [matrix of math nodes,row sep=3em,column sep=4em,minimum width=2em]
  {
     E_{l_\alpha} & D_{l_\alpha} = V(\tau_{l_\alpha}) \\
     & V(\sigma_\alpha) = \cA_\alpha \\};
  \path[-stealth]
    (m-1-1)  edge node [above] {$i_{l_\alpha}$} (m-1-2)
    (m-1-2) edge node [right] {$\pi_{l_\alpha}$} (m-2-2)
            ;
\end{tikzpicture}
\end{equation}
where $ V(\sigma) = \cA_{2,\theta^\ast _\alpha} = \cA_\alpha $ is the two-dimensional base and $ E_{l_\alpha} $ the two-dimensional fiber. The polyhedron for the toric variety $ E_{l_\alpha} $ is nothing but the subpolyhedron of $ \Delta^\ast _4 $ given by $ \theta^\ast _\alpha $ under the projection of $ N $ to $ N(\tau_{l_\alpha}) $ with $ \nu^\ast _{l_\alpha} $ the origin.

The semiample hypersurface $ D^\prime _{l_\alpha} = D_{l_\alpha} \cap Y_4 $ inherits this fibration structure, since the defining polynomial $ p_{\theta_\alpha}=p_\alpha $ is obtained from $ p_\Delta $ by setting all homogeneous coordinates corresponding to integral points in $ \theta^\ast _\alpha $ to zero. This implies in particular that the hypersurface equation is independent of the homogeneous coordinates of $ E_{l_\alpha} $ and therefore, we find the fibration structure
\begin{equation} \label{fiber_D'}
 \begin{tikzpicture}
  \matrix (m) [matrix of math nodes,row sep=3em,column sep=4em,minimum width=2em]
  {
     E_{l_\alpha} & D^\prime_{l_\alpha} = V^\prime (\tau_{l_\alpha}) \\
     & V^\prime (\sigma_\alpha) = R_{\theta_\alpha} \\};
  \path[-stealth]
    (m-1-1)  edge node [above] {$i_{l_\alpha}$} (m-1-2)
    (m-1-2) edge node [right] {$\pi_{l_\alpha}$} (m-2-2)
            ;
\end{tikzpicture}
\end{equation}
where $ V^\prime (\sigma_\alpha) = R_{\theta_\alpha} = R_\alpha $ is the two-semiample hypersurface in $ \cA_\alpha \cap  Y_4 $ defined by the polynomial $ p_\alpha $.

Let us analyze the cohomology of $ D^\prime_{l_\alpha} $. Using the Leray-Hirsch theorem, \cite{hatcher2002algebraic,Voisin03hodgetheory}, we can calculate the cohomology for the fibration
\begin{equation}
 \pi_{l_\alpha} : \quad D^\prime_{l_\alpha} \rightarrow R_\alpha \, 
\end{equation}
with fiber $ E_{l_\alpha} $, that does not degenerate and is locally trivial and the inclusion $ i_{l_\alpha}: E_{l_\alpha} \rightarrow D^\prime_{l_\alpha} $.
For $ c_j \in H^\ast (D^\prime _{l_\alpha}, \mathbb{C}) $ such that $ i_{l_\alpha} ^\ast (c_j) $ generate $ H^\ast(E_{l_\alpha}, \mathbb{C}) $ we find the induced isomorphism of $ \mathbb{C} $-modules
\begin{equation}
 H^\ast(R_\alpha,\mathbb{C}) \otimes_\mathbb{C} H^\ast(E_{l_\alpha},\mathbb{C}) \, \rightarrow \, H^\ast(D_{l_\alpha},\mathbb{C})
\end{equation}
via
\begin{equation}
  b_i \otimes i_{l_\alpha} ^\ast (c_j) \, \mapsto \, \pi_{l_\alpha} ^\ast(b_i) \wedge c_j \, .
\end{equation}
This is not an isomorphism of rings, but makes $ H^\ast(D_{l_\alpha} , \mathbb{C}) $ an $ H^\ast(R_\alpha,\mathbb{C}) $ module. Due to all morphisms appearing here respecting the Hodge structure, the whole isomorphism preserves the Hodge structure. We find therefore, that $ D^\prime _{l_\alpha} $ has Hodge numbers that arise from products of the Hodge numbers of $ R_\alpha $ and $ E_{l_\alpha} $. Here we note that since $ E_{l_\alpha} $ is toric and irreducible, i.e.~connected, its Hodge numbers satisfy
\begin{equation}
 h^{p,q}(E_{l_\alpha}) = 0 \, , \quad p \neq q \, , \quad h^{0,0}(E_{l_\alpha}) = h^{2,2}(E_{l_\alpha}) = 1 \, .
\end{equation}
For the regular semiample hypersurface $ R_\alpha $ of dimension one we find for the independent Hodge-numbers
\begin{equation}
 h^{0,0}(R_\alpha) = 1 \, , \quad h^{1,0}= \ell^\prime (\theta_\alpha) \, .
\end{equation}
 
Recalling \eqref{one_to_three-form}, we hence proved the quality of \eqref{three-cohom}
\begin{equation}
 H^{2,1}( Y_4) \simeq \bigoplus_{\alpha=1}^{n_2} \ \bigoplus_{l_\alpha=1}^{\ell'(\theta_\alpha^*)} H^{1,0}(R_\alpha) \otimes H^{0,0}(E_{l_\alpha})\ ,
\end{equation}
where the first sum runs over all $\theta^\ast _\alpha$ with $\text{dim}(\theta^\ast _\alpha) = 2$ and 
the second sum runs over all $\nu^\ast _{l_\alpha} \in \text{int}(\theta^\ast _\alpha) \cap N$, i.e.~over the divisors $ D^\prime _{l_\alpha} $ that can be blown-down to singular curves $ R_\alpha $.
This can be written using the Hodge star isomorphism as
\begin{equation} \label{five-cohom}
 H^{3,2}( Y_4) \simeq\bigoplus_{\alpha=1}^{n_2} \ \bigoplus_{l_\alpha=1}^{\ell'(\theta_\alpha^*)} H^{1,0}(R_\alpha) \otimes H^{2,2}(E_{l_\alpha}) \, ,
\end{equation}
which also provides a direct match of the form degree.

\section{Picard-Fuchs equations for a toric divisor of $ Y_4 $ with base a Riemann surface} \label{PF_appendix}

With the basics of \autoref{Riemann_periods} introduced, we want to give a detailed description of the period representation of the holomorphic one-forms of Riemann surfaces embedded as hypersurfaces in toric varieties. The trick is to relate the holomorphic forms of the hypersurface to rational holomorphic forms of the ambient space with poles along that hypersurface. These concepts were introduced in \cite{1993alg.geom..6011B} and \cite{1994alg.geom.10017C}, prop 2.1., where we find a general description for the global holomorphic two-forms on $ \cA_2 $ with poles of first order along $ R $ that is a restriction of the anti-canonical hypersurface in $ \cA_5 $
\begin{equation}
 H^0(\cA_2, \Omega^2_{\cA_2}(R)) = \{ \frac{g \, d \omega_{\cA_2}}{p_\theta} \, : g \in S_2 (-K_{\cA_5}|_{\cA_2} + K_{\cA_2}  ) \} \simeq S_2 (-K_{\cA_5}|_{\cA_2} + K_{\cA_2}  ) \, .
\end{equation}
Here $ -K_{\cA_5}|_{\cA_2} \in A_1(\cA_2) $ denotes the Cartier divisor class of the restriction of the anti-canonical divisor of $ \cA_5 $ to $ \cA_2 $ and $ R \subset \cA_2 $ defined by the vanishing of $ p_\theta \in S(-K_{\cA_5}|_{\cA_2}) $. $ -K_{\cA_2} $ is the equivalence class of the anti-canonical divisor of $ \cA_2 $ and also the divisor class of the holomorphic  volume form $ d \omega_{\cA_2} $ which we will discuss below. $ S_2 (-K_{\cA_5}|_{\cA_2} + K_{\cA_2} ) $ denotes the elements of the homogeneous coordinate ring of $ S_2 $ of degree $ [-K_{\cA_5}|_{\cA_2} + K_{\cA_2}]  $. The homogeneous coordinate ring $ S_2 $ of $ \cA_2 $ was discussed after \eqref{S_ndef}.

In the above description of $ H^0(\cA_2, \Omega^2_{\cA_2}(R)) $ appears the holomorphic volume form $ d \omega_{\cA_2} $ on $ \cA_2 $ defined as follows. Consider an index set $ I = \{\nu^\ast _{i^1} , \nu^\ast _{i^2} \}$ consisting of two integral points of $ \Delta^\ast _2 \cap N_2 $. For a fixed integer $ \{ m_1 , m_2 \}$ basis of $ M_2 $ we define
\begin{equation}
 \text{det}(\nu^\ast_I) = \det(\langle m_i, \nu^\ast _j \rangle_{1\leq i,j \leq 2}) \, .
\end{equation}
This enables us to define the holomorphic two-form as
\begin{equation} \label{holomorphic_two_form}
 d \omega_{\cA_2} = \sum_{|I|=2} \text{det}(\nu^\ast _I) \big( \prod_{i \notin I}  X_i \big) dX_{i^1} \wedge dX_{i^2} \, ,
\end{equation}
where the sum runs over all index sets $ I $ with two elements $ \{ i_1 ,i_2 \} $. The grading of this element $ d \omega_{\cA_2} $ is easy to see if we give the differentials $ dX_i $ the same degree as their coordinate counterparts $ X_i $
\begin{equation}
 [\sum_{\nu^\ast _i \in \Delta_2 ^\ast} D_i ] = - K_{\cA_2} \, .
\end{equation}
This enables us to define the Poincar\'e residue as a representation for the holomorphic one-forms of a Riemann surface embedded in a two-dimensional toric ambient space. We can map $ H^0(\cA_2, \Omega^2_{\cA_2}(R)) $ to the holomorphic $(1,0)$-forms of $ R $ by
\begin{align}
 H^0(\cA_2, \Omega^2_{\cA_2}(R)) \quad & \rightarrow \quad H^0(R, \Omega^1 _{R}) \nonumber \\
 \frac{g \, d \omega_{\cA_2}}{p_\theta} \quad & \mapsto \quad \int_{\Gamma} \frac{g \, d \omega_{\cA_2}}{p_\theta}
\end{align}
for $ \Gamma \in H_3(\cA_2 - R,\mathbb{R}) $ a tubular neighborhood of $ R $. Due to partial integration, i.e.
\begin{equation}
 \int_{\Gamma} \frac{g^i \partial_i p_\theta \, d \omega_{\cA_2}}{p_\theta} = 0 \, ,
\end{equation}
it is useful to define the chiral or Jacobian ring for $ p_{\theta} $ as
\begin{equation}
 \cR_\theta = \frac{S_2}{\langle \partial_i p_\theta \rangle} \, ,
\end{equation}
that inherits the grading structure of the homogeneous coordinate ring $ S_2 $ of $ \cA_{2} $. Here $ \langle \partial_i p_\theta \rangle $ denotes the ideal of $ S_2 $ spanned by the partial derivatives of $ p_\theta $.
It was shown in \cite{1993alg.geom..6011B} that this defines an isomorphism
\begin{equation}
 \cR_\theta (-K_{\cA_5}|_{\cA_2} + K_{\cA_2}) \simeq H^{1,0}(R) \, ,
\end{equation}
given by the Poincar\'e residue.

The chiral ring $ \cR_\theta $ can be related to the toric data as follows.
For a divisor $ D_\Delta $ of a toric variety $ \cA $ with polyhedron $ \Delta^\ast $ we have for the degree $ D_\Delta $ submodule $ S([D_\Delta]) $ of the homogeneous coordinate ring $ S $
\begin{equation}
 S([D_\Delta]) = \bigoplus_{\nu \in \Delta} \mathbb{C} \cdot \prod_{\nu^\ast _i \in \Delta^\ast} X_i ^{\langle \nu , \nu^\ast _i \rangle} \, .
\end{equation}
Going to the Jacobian ring $ \cR(p_\Delta) $ for a transverse $ p_\Delta $ reduces the monomials corresponding to vertices and edges of $ \Delta $ to monomials corresponding to points of higher codimension. This implies
\begin{equation}
 \cR_\theta = \bigoplus_{\nu \in \text{int}(\theta)} \mathbb{C} \cdot \prod_{\nu^\ast _i \in \Delta^\ast _2} X_i ^{\langle \nu , \nu^\ast _i \rangle}
\end{equation}
for our example of the Riemann hypersurface $ R $ in $ \cA_2 $ defined by $ p_\theta \in H^0(\cA_2, \cO(K_{\cA_5} |_{\cA_2})) $.

Finally, we can move on to the core topic of our work, the Hodge variation, i.e.~the complex structure dependence of the non-trivial three-forms of a quasi-smooth Calabi-Yau hypersurface $  Y_4 $ in a toric simplicial complete ambient space $ \cA_5 $. As we have seen before, these arise from divisors $ D^\prime _{l_\alpha} $
\begin{equation}
 0 \longrightarrow \bigoplus_{\alpha=1}^{n_2} \bigoplus_{l_\alpha=1}^{\ell'(\theta^*_\alpha)} H^1( D^\prime _{l_\alpha}) \xrightarrow{ \, \oplus \iota_{l_\alpha \ast} \, } H^3( Y_4, \mathbb{C}) \longrightarrow 0 \, ,
\end{equation}
that are two-semiample hypersurfaces of the toric divisors $ D_{l_\alpha} $ of $ \cA_5 $. As discussed before, the full complex structure dependence of a single such divisor is encoded in a Riemann surface $ R $ that is embedded as a hypersurface with equation $ p_{\theta} = 0 $ in the complete simplicial ambient space $ \cA_2 $ with chiral ring $ \cR_\theta $ and holomorphic volume element $ d \omega_{\cA_2} $.

This was already partly analyzed in \cite{Mavlyutov:2000hw} where it was found that we have the isomorphism of $ \mathbb{C} $-modules given by the Poincar\'e residue
\begin{align}
 \cR_\theta (-(1+r)K_{\cA_5}|_{\cA_2} + K_{\cA_2}) &\rightarrow H^{1-r,r}(R) \quad r = 0,1 \nonumber \\
	q	&\mapsto \int_{\Gamma} \frac{q}{p_\theta ^{r+1}} \, d\omega_{\cA_2} \, ,
\end{align}
where $ -K_{\cA_5}|_{\cA_2} $ is the restriction of the anti-canonical divisor defining the fourfold hypersurface and $K_{\cA_2} $ is the canonical divisor of two-dimensional ambient space $ \cA_2 $. The cycle $ \Gamma $ is a tubular neighborhood of $ R_\theta $ in $ \cA_2 $.

The complex structure of our Riemann surface is induced by the complex structure of the ambient Calabi-Yau fourfold whose complex structure we assume to be completely determined by the defining polynomial $ p_\Delta $.
Recall that we consider a family of hypersurfaces of $ \cA_5 $ in the anti-canonical class $ K_{\cA_5} $ given by the family of polynomials, as already described in \eqref{p_Delta},
\begin{equation}
p_\Delta (a) = \sum_{\nu_j \in \Delta} a_j \prod_{\nu^\ast _i \in \Delta^\ast} X_i ^{\langle \nu_j , \nu^\ast _i \rangle + 1} \quad \in S(- K_{\cA_5}) \, .
\end{equation}
The complex structure deformations, and we consider for simplicity only the algebraic deformations by monomials
\footnote{A toric divisor $ D^\prime _2 $ with holomorphic two-forms induces non-algebraic complex structure deformations of $ Y_4 $. If $ D^\prime _1 $ carrying holomorphic one-forms intersects $ D^\prime _2 $, the induced three-forms on $ Y_4 $ can depend on these non-algebraic complex structure deformations. This can be investigated using a realization of $ Y_4 $ as complete intersection with all complex structure deformations algebraic.}, for this hypersurface are given by
\begin{equation}
H^{3,1}(Y_4)_{alg} \simeq \cR(p_\Delta) (- K_{\cA_5}) = \frac{\mathbb{C}[\prod_{\nu^\ast _i \in \Delta^\ast} X_i ^{\langle \nu , \nu^\ast _i \rangle + 1}]}{\langle \partial_i p_\Delta \rangle}
\end{equation}
which can be represented by all monomials $ p_\nu $ for $ \nu \in \Delta \cap M $ that is not a vertex or part of an edge of $ \Delta $, i.e.~does not lie in the interior of a face of dimension less than two.

Since the complex structure of the Riemann surface at the complex structure point $ a $, denoted by $ (R)_a $, is induced by the complex structure of the fourfold at $ a $, denoted by $ (Y_4)_a $, the monomial complex structure deformations of $ (R)_a $ are represented by the monomials corresponding to the interior points of $ \theta $, i.e.~by $ \cR_\theta (-K_{\cA_5}|_{\cA_2}) $. Therefore, we find for $ p_j \in S(-K_{\cA_5}) $ a monomial variation corresponding to an integral point $ \nu_j \in \Delta - \text{int}(\theta) $ that
\begin{equation}
\frac{\partial}{\partial a_j} \, \gamma_{b}(a) = 0 \, , \quad \forall \gamma_{b} \in H^1((R)_a, \mathbb{C}), \, \nu_j \notin \text{int}(\theta) \, .
\end{equation}
This justifies to denote the complex structure coordinates on which the complex structure of $ R $ depends, i.e.~the polynomial $ p_\theta $ depends, as $ a_{\nu_b} = a_{b} $, since we denoted by $ \nu_b $ the integral points contained in the interior of $ \theta $. Note here also that the holomorphic one-forms $ \gamma_c (a) $ depend holomorphically on the complex structure moduli $ a_b $, which also implies that the normalized period matrix $ \hat f_{ab}(a) $ is a holomorphic function of the complex structure coordinates $ a $.

Using the residue expressions as local trivialization of the Hodge bundles with fibers in $ H^1((R)_a,\mathbb{C}) $ over complex structure moduli space, we can derive the complex structure dependence of the $ (1,0) $-forms
\begin{equation}
 \gamma_{b} (a) = \int_{\Gamma} \frac{p^\prime _b}{p_\theta (a)} \, d\omega_{\cA_2} \, , \in H^{1,0}((R)_{a}) , \quad \nu_{b} \in \text{int}(\theta) \cap M\, .
\end{equation}
with $ p^\prime _b = p_{\nu_b} / \prod_{\nu^\ast _i \in \theta^\ast} X_i $.
Taking a simple partial derivative leads to
\begin{equation}
 \frac{\partial}{\partial a_c} \, \gamma_b (a) = \frac{\partial}{\partial a_b} \, \gamma_c (a) = -\int_{\Gamma} \frac{p^\prime _b p_c}{p_\theta ^2(a)} \, d\omega_{\cA_2} \,, \quad \in H^1((R)_a,\mathbb{C}) \, ,
\end{equation}
where
\begin{equation}
 \frac{\partial}{\partial a_c} \, \gamma_b (a) \in H^{1,0}((R)_a) \quad \text{for} \;  p_b ^\prime p_c \in \langle \partial_i p_\theta \rangle \, ,
\end{equation}
and
\begin{equation}
 \frac{\partial}{\partial a_c} \, \gamma_b (a) \in H^{0,1}((R)_a) \quad \text{for} \;  p_b ^\prime p_c \notin \langle \partial_i p_\theta \rangle \, .
\end{equation}

Since this already exhausts the one-dimenisonal cohomology groups, we find that
\begin{equation}
 \frac{\partial}{\partial a_c} \frac{\partial}{\partial a_d} \, \gamma_b (a) = 2 \int_{\Gamma} \frac{p^\prime _b p_c p_d}{p_{\theta} ^3(a)} \, d\omega_{\cA_2}
\end{equation}
and hence that for degree reasons
\begin{equation}
 p^\prime _b p_c p_d \in \langle \partial_i p_\theta \rangle \, .
\end{equation}
From this we can deduce that the second derivative of a holomorphic one-form $ \gamma_b(a) $ can be expressed as a linear combination of the $ \gamma_b(a) $ and its first derivatives with coefficients rational functions of the complex structure moduli $ a_c $.
In practice, we can express the second derivatives of $ \gamma_b $ by operators acting on $ \gamma_b $ of the form
\begin{equation}
\frac{\partial}{\partial a_c} \frac{\partial}{\partial a_d} \, \gamma_b (a) = \big( c^{(1)}{(a)_{c d b e}}^ f \frac{\partial}{\partial a_e} + c^{(0)}{(a) _{c d b}}^f \big) \, \gamma_f (a) \,,
\end{equation}
where $ c^{(1)}(a) _{c d b e} \, ^ f$, $c^{(0)}(a) _{c d b} \, ^f $ are rational functions of the complex structure moduli $ a_c $ that are completely symmetric in their lower four, respectively three, indices. These functions are structure constants of the chiral ring $ \cR_\theta $ determining the multiplication rules in this ring.
The above differential relations are called Picard-Fuchs equations and can be used to determine the complex structure dependence of the holomorphic one-forms on $ R $. In particular, we note that
\begin{equation}
 \frac{\partial}{\partial a_c} \, \gamma_b (a) = \frac{\partial}{\partial a_b} \, \gamma_c (a)
\end{equation}
is an integrability condition, allowing us to find a one-form valued prepotential $ \gamma(a) $ that satisfies $\gamma_b = \frac{\partial}{\partial a_b} \gamma$.
It is suggestive that the structure constants  $ c^{(1)}{(a)_{c d b e}}^ f$, $c^{(0)}{(a)_{c d b}}^f $ are the same structure constants that arise from the whole chiral ring $ \cR(p_\Delta) = \cR $ from which $ \cR_\theta $ is constructed as a quotient.

\bibliography{bibmaster}
\bibliographystyle{utcaps}

\end{document}